\documentclass[lettersize,journal]{IEEEtran}
\usepackage{amsmath,amsfonts}
\usepackage{algorithmic}
\usepackage{algorithm}
\usepackage{array}
\usepackage[caption=false,font=normalsize,labelfont=sf,textfont=sf]{subfig}
\usepackage{textcomp}
\usepackage{stfloats}
\usepackage{url}
\usepackage{verbatim}
\usepackage{graphicx}
\usepackage{cite}
\usepackage{soul}
\usepackage{multirow}
\usepackage{longtable}
\usepackage{tikz} 
\usepackage{adjustbox}
\usepackage{makecell}
\usepackage{balance}
\usepackage{orcidlink}

\newtheorem{definition}{Definition}
\newcolumntype{?}{!{\vrule width 1.5pt}}
\newcolumntype{P}[1]{>{\centering\arraybackslash}p{#1}}

\begin{document}

\title{Eye-tracked Virtual Reality: A Comprehensive Survey on Methods and Privacy Challenges}

\author{Efe Bozkir, Süleyman Özdel, Mengdi Wang, Brendan David-John, Hong Gao, Kevin Butler, Eakta Jain, and Enkelejda Kasneci
\thanks{S. Özdel and M. Wang contributed equally.}
\thanks{E. Bozkir, S. Özdel, M. Wang, and E. Kasneci are with Technical University of Munich, School of Social Sciences and Technology, Department of Educational Sciences, Chair for Human-Centered Technologies for Learning, Germany (E-mails: \{efe.bozkir, ozdelsuleyman, mengdi.wang, enkelejda.kasneci\}@tum.de).}
\thanks{H. Gao is with Soochow University, China (E-mail: gaohong@suda.edu.cn).}
\thanks{B. David-John is with Virginia Tech, USA (E-mail: bmdj@vt.edu).}
\thanks{K. Butler and E. Jain are with the University of Florida, USA (E-mails: butler@ufl.edu, ejain@ufl.edu).}
}



\maketitle

\begin{abstract}
The latest developments in computer hardware, sensor technologies, and artificial intelligence can make virtual reality (VR) and virtual spaces an important part of human everyday life. Eye tracking offers not only a hands-free way of interaction but also the possibility of a deeper understanding of human visual attention and cognitive processes in VR. Despite these possibilities, eye-tracking data also reveals users' privacy-sensitive attributes when combined with the information about the presented stimulus. To address all, this survey first covers major works in eye tracking, VR, and privacy areas between 2012 and 2022. While eye tracking in VR part covers the computational eye tracking pipeline from pupil detection and gaze estimation to offline data analysis, for privacy and security, we focus on eye-based authentication as well as computational methods to preserve the privacy of individuals and their eye-tracking data in VR. Later, we outline three main directions by focusing on privacy. In summary, this survey presents an extensive literature review of the utmost possibilities of eye tracking in VR and their privacy implications. 
\end{abstract}


\section{Introduction}
Over the last decade, virtual, augmented, and mixed reality (VR/AR/MR) communities have benefited from advancements in computer hardware, graphics, and imaging science. To date, some modern head-mounted displays (HMDs) have become available at reasonable prices for everyday use. Furthermore, eye-tracking sensors have become either directly integrated into these HMDs (e.g., HTC Vive Pro Eye~\cite{vrcompare_htcviveproeye} or are available as low-cost add-ons (e.g., Pupil Labs~\cite{pupillabs_addon_opensource_paper}). Since eye movements are related to human cognition, visual attention, and perception, the information on where users look can be beneficial to them in various ways, such as providing adaptive support in VR, increasing their engagement, or enhancing the usability of VR applications. At the same time, there is also considerable room for technical improvements, such as enhancing display resolutions, increasing realism, or preventing cybersickness, in which gaze information may be helpful. 

It was foreseen in 2017 that VR was approximately 20 years away from achieving resolutions comparable to those of the human eye~\cite{nvidia_estimate_VR}. With such advancements and a wider range of applications, future HMDs may evolve into mobile devices, similar to today's mobile phones and tablets. However, as HMDs are situated on the users' heads and have close proximity to their eyes, they may be perceived as privacy-invasive, as many sensitive attributes about users, such as gender, sexual preferences, and identities, can be extracted using eye movements~\cite{Liebling2014}. From a privacy perspective, this information should be protected, or the user should have control over its release. At the same time, a pleasant virtual experience should be available to all users. Furthermore, eye information can also be utilized for accurate authentication, especially with iris textures, as they are analogous to fingerprints. Therefore, there is an essential need for a privacy-utility trade-off.  

As VR and privacy communities have been working to tackle these issues, and with the availability of HMDs to wider communities, especially in recent years, the number of works at the intersection of VR, eye tracking, and privacy has been increasing. However, to date, none of the previous works have focused on comprehensive coverage of eye tracking in VR and its privacy implications, considering the leading venues in the field. In this work, \textbf{(1)} we first cover the research between 2012 and 2022 by discussing how eye movements are extracted and the possibilities with them in immersive VR. Then, \textbf{(2)} we discuss security and privacy implications, including eye-based authentication and privacy-preserving eye tracking. Lastly, \textbf{(3)} we draw and discuss three different directions for future research, especially focusing on privacy. 

\subsection{Paper's Aim, Structure, and a Navigational Guide}
This paper aims to provide researchers and practitioners with a detailed overview of the different aspects of eye tracking in VR. We begin with details on existing hardware, datasets, and algorithms, the latter being more specific to eye-region segmentation and gaze estimation. We then dig deeper into eye-based interaction techniques in VR, which often inherently utilize estimated eye regions and gaze. Furthermore, the overview of works on human cognition, visual attention, and perception in VR aims to teach the reader about the relationship between human eye movements and physiology. The security and privacy aspects discussed in the paper aim to provide methodological details on the relationship between eye movements and user authentication, and how to protect users' privacy when the focus is not on authentication. Lastly, future directions involve a privacy-aware approach to eye movement-based methods in VR. Overall, the primary objective of this paper is to provide a comprehensive guide to navigating the landscape of eye tracking in VR, encompassing its technical foundations, practical applications, and ethical considerations, while also suggesting directions for further exploration, particularly in the context of privacy. 

As we comprehensively survey research conducted in eye tracking, VR, and privacy domains, and due to the interdisciplinary nature of each domain, we provide a navigational guideline on how readers with different backgrounds can benefit from it. We first discuss existing works in the literature in Section~\ref{sec_rw}. We then discuss our search methodology, considering the most important academic venues in Section~\ref{sec_method}. These sections are suitable for all readers, particularly those seeking to understand existing surveys and this work's novelty.  

Section~\ref{sec_eyetracked_vr} discusses a range of topics, from computer vision-based approaches for tracking eyes and further processing eye region data to eye-based human-computer interaction (HCI) and understanding human visual attention, cognition, and perception. Section~\ref{sec_4_1} will benefit most readers interested in computer vision and machine learning. Section~\ref{sec_4_2} focuses on HCI by utilizing eye movements for various purposes, including real-time interaction in VR, foveated rendering, and addressing technical issues to enhance immersive experiences, which is suitable for readers who work on the technical aspects of HCI and eye tracking. Section~\ref{sec_4_3} combines eye tracking and VR from a human cognition, attention, and perception perspective, and is suitable for those with cognitive and experimental psychology backgrounds, or those interested in conducting human factors research. 

\begin{figure*}[t]
  \centering
  \includegraphics[width = 0.98\linewidth]{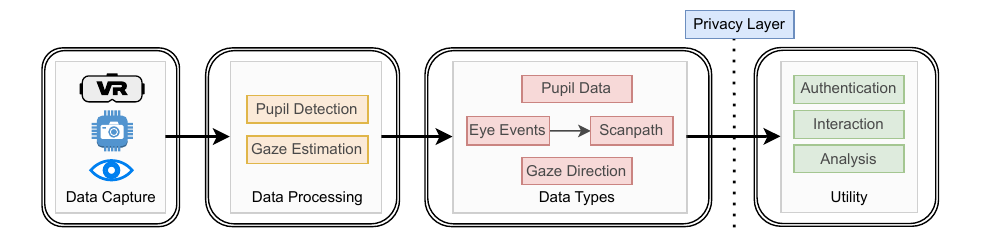}
  \caption{Visual representation of eye-tracked VR, considering data capturing, processing, data types, and utility tasks.}
  \label{VR_eye_tracking_processing_diagram}
\end{figure*}

Section~\ref{sec_privacy} presents works exploring the privacy and security implications of eye-tracked VR, including authentication possibilities, the importance of privacy-preserving methods, and how current literature has addressed privacy issues, which is relevant for both privacy and security researchers in HCI. Lastly, although we primarily discuss the future directions and implications of eye-tracked VR in Section~\ref{sec_discussion} with a focus on privacy, this section is relevant for all researchers whose work involves VR and eye tracking, as privacy issues should be carefully handled regardless of the research area. Section~\ref{sec_conc} concludes our paper. In summary, after reading this paper in whole or in part, we anticipate that it will be easier for novice individuals to become knowledgeable about the state-of-the-art, while our paper serves as a concise reference for experienced researchers. Figure~\ref{VR_eye_tracking_processing_diagram} provides a holistic visual representation of the concepts discussed in this paper. 

\section{Related work}
\label{sec_rw}
Several works analyze relevant research for eye tracking, VR/AR, privacy, and security. None systematically and comprehensively reviews eye-tracking in VR setups by covering the major venues, the privacy issues that such works could lead to, and how to mitigate such risks. Previously, \cite{Lappi_2015_good_bad_ugly} analyzed eye tracking in the wild by discussing the advantages, disadvantages, and technical terms to achieve high-quality results from eye-tracking experiments. Furthermore, \cite{Plopski_etal_2022_CSUR} covered gaze-based interaction and eye tracking in head-worn extended reality (XR). Similarly, \cite{eye_track_support_for_Vis_Analytics_inc_privacy_ETRA19} discussed the foundations of eye-tracking support for visual analytics systems, applications, and challenges by five critical themes, including privacy protection as an essential issue. 

Previous work has studied the challenges in immersive analytics by considering VR, MR, and AR with future directions~\cite{grand_challenges_in_immersive_analytics_chi21}, sector- and domain-specific reviews, such as for sports, highlighting the importance of usability, interaction, and image presentation~\cite{Pastel_etal_2023}. Furthermore, \cite{eyetracking_in_vr_jemr} examined eye-tracking research in VR, with a pilot user study that focused on experimental setups and common problems, such as vergence-accommodation conflict and eye-tracking calibration, on a small scale. \cite{adhanom2023review} broadly reviewed eye-tracking applications in VR, and similar to~\cite{eye_track_support_for_Vis_Analytics_inc_privacy_ETRA19}, they identified privacy and security as essential discussion points. However, their review was not comprehensive and covered privacy as an application with a handful of works. A similar review focused on eye-tracking processing is given in~\cite{Moreno_Arjonilla2024}, and~\cite{10087032} briefly overviews the privacy aspects. Furthermore, overviews on high-quality eye-tracking data collection~\cite{considerations_for_VR_eyetracking_etra21} and the measurement of data quality in HMD-based eye tracking~\cite{b2020gazemetrics} also exist. 

While visual scanning patterns and eye-tracking data provide useful information about users, this information also raises concerns about users' privacy. \cite{Liebling2014} discussed that the end-users may not understand the privacy loss and argued that privacy protection should be in place for gaze and pupillometry. \cite{Kroeger2020} highlighted the attributes that can be inferred from gaze data, similar to~\cite{Liebling2014}, and discussed the privacy implications and societal impacts. \cite{sec_privacy_for_AR_SP18} considered multi-user AR's security and privacy and found that some users were concerned about the powerful abilities of the eye-tracking enabled interfaces and behavioral tracking, such as an HMD understanding that the user is being attracted to someone due to being unable to stop looking at them. This finding partly aligns with the findings of more recent work~\cite{steil_dp_etra19}, which indicate that users are more likely to agree to share their eye-tracking data if the data owner is a governmental health agency or if the purpose is research. Furthermore,~\cite{katsini_secpriv_survey_chi20} provided an extensive survey on gaze-based authentication, encouraging further research on privacy-preserving eye tracking like prior works~\cite{adhanom2023review, eye_track_support_for_Vis_Analytics_inc_privacy_ETRA19}.  

Despite several recent attempts, none of the previous works have comprehensively and systematically analyzed the research on eye-tracking data analytics in VR, considering the entire processing pipeline of eye-tracking methods. We consider methods from pupil detection and gaze estimation to human visual attention and cognition for processing and mining eye-tracking data. We further explore works that investigate privacy-preserving methods for eye movements in VR setups, providing novel research directions that should be pursued to enhance the practical applications of VR. Considering the number of sensitive attributes that eye-tracking data can reveal, we take privacy as one of the main pillars of our paper. Therefore, our survey, focusing on these aspects, provides a comprehensive overview for the research community. 

\section{Methodology}
\label{sec_method}
We first analyzed the recent surveys and papers that study eye tracking in the context of VR, AR, HCI, and privacy~\cite{Lappi_2015_good_bad_ugly,eye_track_support_for_Vis_Analytics_inc_privacy_ETRA19,grand_challenges_in_immersive_analytics_chi21,Plopski_etal_2022_CSUR,mr_ar_overview_ismar20,Liebling2014,Kroeger2020,katsini_secpriv_survey_chi20}. We considered the papers between 2012-2022 since the VR devices and eye trackers have been becoming prevalent lately. We used Google Scholar to query papers from 35 renowned venues including ACM CHI, IEEE VR, ACM ETRA, IEEE ISMAR, ACM IUI, ACM UIST, ACM ICMI, ACM MobileHCI, ACM MM, ACM VRST, IEEE AIVR, AAAI Conference on Artificial Intelligence, IEEE CVPR, IEEE ICCV, ECCV, NeurIPS/NIPS, ACM SUI, IEEE VIS, PoPETS/PETS, USENIX Security, USENIX SOUPS, NDSS, IEEE Symposium on S\&P, ACM CCS, ACM Interactive, Mobile, Wearable and Ubiquitous Technologies (IMWUT), Journal of Eye Movement Research (JEMR), IEEE Transactions on Visualization and Computer Graphics (TVCG), ACM Transactions on Graphics (ToG), IEEE S\&P Magazine, Journal of Privacy and Confidentiality, IEEE Transactions on Information Forensics and Security (TIFS), IEEE Transactions on Dependable and Secure Computing, Computers \& Security (C\&S), ACM Transactions on Privacy and Security, and ACM Transactions on Cyber-Physical Systems. For the conference proceedings, we considered the main proceedings, excluding companion and workshop proceedings, except for works that present concrete results or are highly related to eye tracking and VR. Proceedings of the ACM ETRA in 2022 were published in the Proceedings of the ACM on Human-Computer Interaction and the ACM on Computer Graphics and Interactive Techniques, and we also included those in our search results under the ETRA category. We cross-checked the query results from the DBLP. 

\begin{figure}[h]
  \centering
  \includegraphics[width = \linewidth]{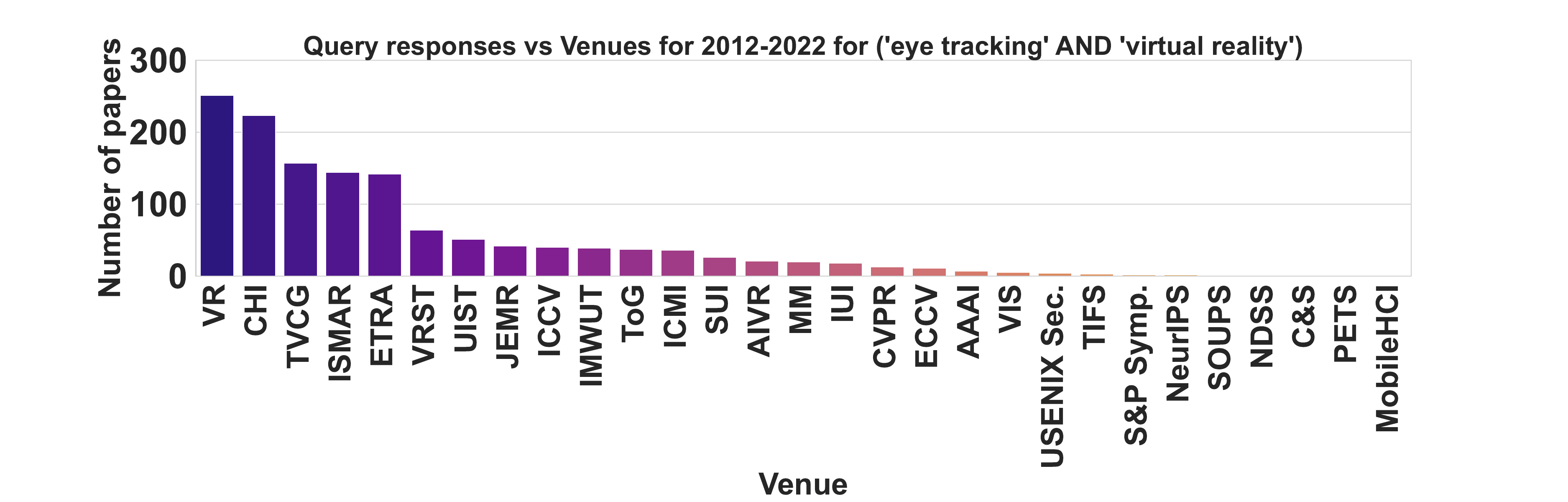}
  \caption{Query responses according to venues for our main query.}
  \label{fig:main_query_distribution}
\end{figure}

To filter the papers, we used (``eye tracking'' AND ``virtual reality'') as the main query, where each phrase could appear anywhere in the papers. In addition, we used the following sub-queries to further track privacy- and biometrics-related papers: (``eye tracking'' AND ``virtual reality'' AND  ``privacy'') and (``eye tracking'' AND ``virtual reality'' AND  ``biometrics''). We also applied forward and backward tracking by reviewing the citations and references of the queried papers, respectively. The main query yielded 1364 results, whereas two sub-queries led to 215 and 96 results, respectively. Figure~\ref{fig:main_query_distribution} shows the resulting distributions of the main query. If a venue above does not appear in Figure~\ref{fig:main_query_distribution}, it means that we did not find any paper at this venue with our query. 

\section{Eye-tracked Virtual Reality}
\label{sec_eyetracked_vr}
Eye tracking can be considered one of the key technologies, especially for VR, as it enables the accurate tracking of attention within HMDs. In addition, since eyes do not move completely voluntarily, one can obtain more objective user behaviors compared to self-reported data, such as questionnaires. While one can utilize this rich source of information in various ways, there are different technical steps to achieve this. Firstly, eye regions and gaze directions are extracted primarily using computer vision and machine learning techniques. Later, gaze directions are utilized for interactions, for instance, to support users by predicting their intents and providing them with context-sensitive aid or technical ways, such as foveated rendering. While such cases should work in real-time, collected data from different experimental setups can also be used offline to understand humans and how they behave in various application domains, such as education or medicine, showing that VR can also be considered a research tool to create scenarios that are almost impossible in the wild due to privacy and safety issues or because of practical infeasibility. 

Considering all, we organize this section as follows. We first discuss works that utilize computer vision and machine learning to estimate gaze and detect eye regions, as these are the initial steps in data processing, with a significant portion of recent research on these topics employing deep learning-based approaches. We also include eye movement event detection methods in VR, which differ from those in conventional settings. We then provide a detailed overview of eye-based interaction and its various applications, with a particular focus on efficiency, usability, and real-time working capability. As the last step, we cluster the works that conduct offline data analyses, particularly to understand human visual attention, cognition, and perception, and we provide an overarching view of these works. While the offline understanding of human visual attention, cognition, and perception does not have a direct and real-time impact on the interaction experience, understanding these aspects helps to achieve an optimal virtual space and interaction design. 

\subsection{Hardware, Datasets, and Algorithms}
\label{sec_4_1}
We first provide insights into the works on hardware and datasets related to eye-tracked VR in Sections~\ref{subsubsec:hardware} and~\ref{subsubsec:dataset}, respectively. Then, in Sections~\ref{subsubsec:segmentation},~\ref{subsubsec:estimation}, and~\ref{subsubsec:event}, we discuss algorithms for eye region segmentation, gaze estimation, and eye event classification, respectively. Table~\ref{table:sec4.1} briefly summarizes the papers in this section. Eye tracking in immersive VR using HMDs is fundamentally different from real-world scenarios, as users typically wear HMDs on their heads, and sensors detect the eye regions rather than the entire face. Therefore, unless explicitly mentioned, we focus on such setups. For an extended evaluation of eye tracking, such as gaze estimation, regardless of immersive VR or the real world, we refer the reader to the surveys on appearance-based gaze estimation with deep learning~\cite{appearance_based_gaze_est_dl_arxiv2021}, convolutional neural network (CNN)-based gaze estimation~\cite{CNN_based_gaze_est_survey_2020}, and gaze estimation in consumer platforms~\cite{8003267}. 

\begin{table}[h!]
\scriptsize
\centering
\caption{\label{table:sec4.1} A broad overview of papers in Section~\ref{sec_4_1}.}
\begin{tabular}{ll}
Contribution & Characteristics \\ 
\hline \hline
\multirow{3}{*}{\begin{tabular}[l]{@{}l@{}} Eye-tracker \\ performance \\ assessment \end{tabular}} & \cite{b2020gazemetrics, schuetz2022eye} HTC Vive Pro Eye \\ 
 & \cite{lohr2019evaluating, Roth_Weier_Hinkenjann_Li_Slusallek_2017} SMI add-on eye tracker + HTC Vive / Oculus Rift \\ 
 &\cite{stein2021comparison} HTC Vive Pro Eye, Varjo VR-1, Fove-0 \\ 
\hline
 \multirow{4}{*}{\begin{tabular}[l]{@{}l@{}} Low-cost \\ eye trackers \end{tabular}} & \cite{Stengel2015cheap} Using dichroic mirrors and personalizable lens \\ 
 &\cite{low_cost_gaze_tracking_google_cardboard_vrst16} Using phone selfie camera + Google Cardboard \\ 
 &\cite{ahuja_eyespyVR_lowcost_IMWUT18} EyeSpyVR: Using phone selfie camera + VR Box headset \\ 
 &\cite{EyeMR_lowcost_prototyping_Cardboard_ETRA18} EyeMR: Using USB camera + IR-LED + Cardboard \\ 
\hline
 \multirow{3}{*}{\begin{tabular}[l]{@{}l@{}} Non-VOG \\ eye trackers \end{tabular}} &\cite{optical_gaze_tracking_ismar20, power_efficient_ET_sensor_for_portable_VR_headsets_ETRA19} Photosensor-oculography (PSOG) \\ 
 &\cite{shimizu2016eye, bernal2022galea} Electro-oculography (EOG) \\
 &\cite{whitmire2016eyecontact} Scleral search coil (SSC) \\
\hline
 \multirow{12}{*}{Datasets} &\cite{OpenEDS_Open_Eye_Dataset, garbin_eyetrackingdataset_etra20} OpenEDS: 152 users, for eye region segmentation \\ 
 &\cite{NVGaze_chi19} NVGaze: 35 users, synthetic + real data \\  
&\cite{Jiang_2020_CVPR} IQVA: 14 users, question-driven visual attention \\ 
&\cite{360_event_detection_Agtzidis_MM_19} 13 users, automatic eye event classification \\
 &\cite{fuhl2021TEyeD} TEyeD: 54 users, largest unified eye dataset \\
&\cite{jin2022you} 100 users, quantitative taxonomy for videos, multi-modal \\ 
 &\cite{OpenNEEDS_dataset_etra21} OpenNEEDS: 44 users, multi-modal \\ 
 &\cite{tabbaa2022VREED} VREED: 34 users, for emotion recognition, multi-modal \\
 &\cite{zhang2022egobody} EgoBody: 2 users, for human body reconstruction \\
&\cite{RCEA-360VR_emotion_annotation_in360vids_chi21}  32 users, real-time emotion annotation \\  
\hline
 \multirow{7}{*}{\begin{tabular}[l]{@{}l@{}} Eye region \\ segmentation \end{tabular}} &\cite{eyenet_segmentation_ICCVW19} EyeNet: Using residual blocks and convolutional attention \\ 
 &\cite{Fuhl_2019_ICCV} Using CycleGANs for segmentation and generation \\ 
 &\cite{ellseg_semantic_seg_for_gaze_tracking_TVCG21} EllSeg: Robust against occlusions \\ 
 &\cite{domain_adaptation_for_eye_segmentation_ECCVW20}  Semi and unsupervised domain adaptation \\ 
 &\cite{Boutros_2019_ICCV} Eye-MMS: Using multi-scale inter-connected CNN \\ 
 &\cite{eyeseg_ECCVW20} EyeSeg: Using generalized dice loss function \\ 
 &\cite{chaudhary_ritnet_iccvw_19} RITnet: Using U-Net + DenseNet\\ 
\hline
 \multirow{3}{*}{\begin{tabular}[l]{@{}l@{}} Generating \\ eye image from \\ segmentation \end{tabular}} &\cite{D-ID-Net_identity_preserving_image_generation_iccvw19}   D-ID-Net: Two-phase image generation \\ 
&\cite{Buehler2019Seg2Eye} Seg2Eye: content from segmentation + style from person \\ 
 &\cite{lu2022geometry} GeoMaskGAN: Maintaining geometric consistency \\
\hline
 \multirow{7}{*}{\begin{tabular}[l]{@{}l@{}} \\ Gaze \\ estimation \end{tabular}} &\cite{feng2022real} Using changes in pixel brightness, lightweight \\
 &\cite{beyond_10khz_paper_TVCG2021} Using eye motion event, extremely high frame rate \\
 &\cite{lu2022neural} Neural3DGaze: 3D pupil localization \\
 &\cite{Ranjan_2018_CVPR_Workshops} Robust against head pose changes \\ 
 &\cite{Yu_2020_CVPR} Using unsupervised representation + gaze redirection \\ 
 &\cite{benefit_of_temporal_info_for_appearace_based_gaze_est_etra20} Using CNN-recurrent model for temporal gaze trace \\ 
&\cite{Cheng_2018_ECCV} ARE-Net: Asymmetric regression of eyes \\ 
&\cite{stojanov2022benefits} Using depth information \\ 
\hline
 Calibration &\cite{gaze_beh_interacted_obj_VR_ETRA19} Using correlation between fixation and hand interaction \\
\hline
 \multirow{3}{*}{\begin{tabular}[l]{@{}l@{}} Eye event \\ classification \end{tabular}} &\cite{EM_classification_HMD_360_agtzidis_and_dorr_ETRA19} Data/algorithm translation between 2D-monitor and HMD \\ 
 &\cite{360_event_detection_Agtzidis_MM_19} Rule-based classifier \\ 
&\cite{rolff2022saccades} Estimation of remaining time until next saccade \\ 
\hline
\end{tabular}
\end{table}

\subsubsection{Hardware} 
\label{subsubsec:hardware}
\begin{figure}[t]
  \centering
  \includegraphics[width = 1\linewidth]{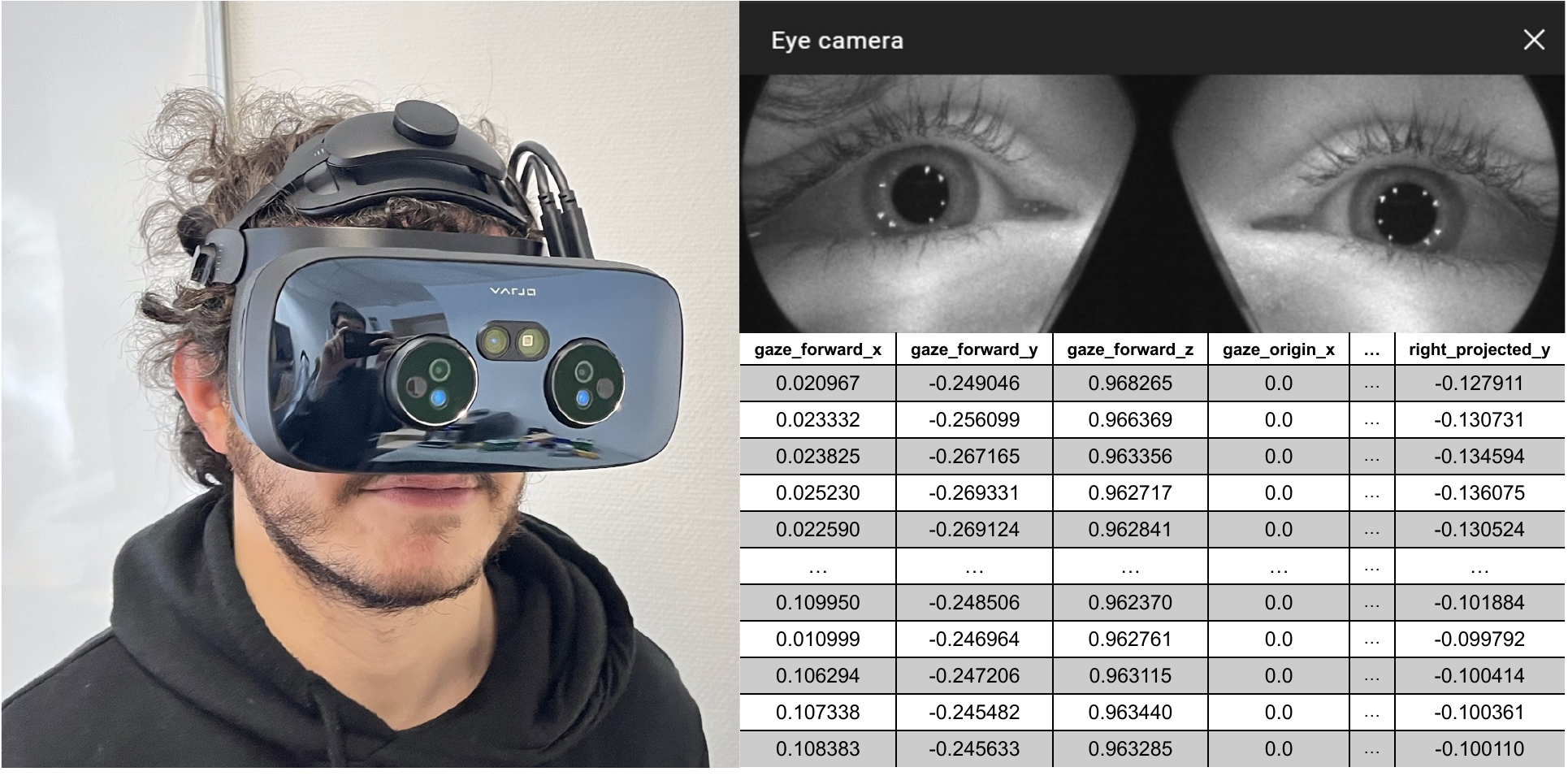}
  \caption[xr3]{Left: Varjo XR-3, a high-end HMD device for XR. Upper right: eye images taken within the HMD. Lower right: some eye features recorded.}
  \label{fig:varjo_xr3}
\end{figure}

Most VR headsets (e.g., Oculus Rift~\cite{vrcompare_oculusrift}) formerly did not deliver built-in eye trackers. Practitioners needed add-on eye trackers (e.g., Pupil Labs~\cite{pupillabs}, SensoMotoric Instruments (SMI), acquired by Apple~\cite{smi_techcrunch}) to enable eye tracking functionality. As eye tracking and VR grow at an astounding rate, an increasing number of HMDs like HTC Vive Pro Eye~\cite{vrcompare_htcviveproeye} and Varjo XR-3~\cite{varjo_xr3_link} support eye tracking on their own. Figure~\ref{fig:varjo_xr3} illustrates an example of an HMD, the Varjo XR-3, which features an integrated eye tracker, with eye images and a portion of sample raw data. Meanwhile, a growing body of research highlights the importance of evaluating the performance of built-in and add-on eye trackers for HMDs. For instance,~\cite{lohr2019evaluating} measured data quality acquired with the SMI eye tracker in the HTC Vive from multiple facets, including spatial accuracy and precision, temporal precision, linearity, and crosstalk. Eye-tracking data quality from an SMI extension has also been analyzed~\cite{Roth_Weier_Hinkenjann_Li_Slusallek_2017}, considering tracking precision and fixation accuracy in a foveated rendering context.

An average eye-tracking accuracy of 1.23$^{\circ}$ and a root mean square (RMS) precision of 0.62$^{\circ}$ were also reported on the HTC Vive Pro Eye with a Unity package called GazeMetrics~\cite{b2020gazemetrics}. In a recent work~\cite{schuetz2022eye}, an average accuracy of 1.08$^{\circ}$ and mean standard deviation (SD)/RMS precisions of 0.36$^{\circ}$/0.2$^{\circ}$ were reported on the same device after outlier correction. The authors also found a significant decrease in accuracy and precision when participants wore vision correction glasses, while the effect of contact lenses was more elusive. \cite{stein2021comparison} conducted a comparison between Fove-0, Varjo VR-1, and HTC Vive Pro Eye using two other metrics: eye-tracking delay and latency. With a delay of 15-52 ms and a latency of 45-81 ms, Fove-0 outperformed the other HMDs. 

\paragraph{Low-cost HMD Eye Trackers}
While add-on or built-in eye trackers have become more affordable, they can still be expensive due to hardware and software requirements. To address this,~\cite{Stengel2015cheap} proposed an implementation utilizing dichroic mirrors and a personalizable lens positioning system at an estimated cost of \$450. The authors deployed dichroic mirrors to support eye-tracking cameras outside of the field of view (FOV), whereas the lens locating module accounts for different interocular distances. With a model-based gaze estimation algorithm, an angular error of 0.5$^{\circ}$-3.5$^{\circ}$ was achieved. 

An affordable, alternative system that utilizes a smartphone and a Google Cardboard~\cite{google_cardboard}, costing approximately \$15, also exists~\cite{low_cost_gaze_tracking_google_cardboard_vrst16}. In this system, gaze estimation is achieved by capturing Purkinje images (i.e., reflections of on-screen images on the eyes) using the smartphone's front camera, resulting in an average error of 5.0$^{\circ}$. In parallel,~\cite{ahuja_eyespyVR_lowcost_IMWUT18} presented a similar framework named EyeSpyVR, which also uses the smartphone selfie camera and a \$10 VR Box headset to achieve coarse gaze tracking. With 70 subjects, the authors reported an average gaze estimation error of 10.8$^{\circ}$ and 12.9$^{\circ}$, respectively, with and without calibration. \cite{EyeMR_lowcost_prototyping_Cardboard_ETRA18} deployed another system on Google Cardboard called EyeMR. In contrast to systems realized using smartphone selfie cameras, EyeMR utilizes USB cameras and extended IR-LED circuit boards to perform gaze estimation, supporting both monocular and binocular eye tracking. 

\paragraph{Non-Video-oculography (VOG) HMD Eye Trackers}
Standard camera-based eye-tracking systems fall into the category of video-oculography (VOG). VOG-based methods often struggle to strike an optimal balance among factors such as power consumption, computational cost, latency, and accuracy. Other types of eye trackers also exist, which can be categorized into photosensor-oculography (PSOG), electro-oculography (EOG), and scleral search coil (SSC)~\cite{duchowski2017eye, rigas2018photosensor}. PSOG is akin to VOG because they both often require light sources and optical sensors to capture reflections. In contrast, PSOG usually uses a sparse grid of photodiodes and photosensors, recording significantly fewer pixels than VOG. As a result, PSOG can outperform VOG in terms of sampling rate, computational cost, and power consumption~\cite{optical_gaze_tracking_ismar20, power_efficient_ET_sensor_for_portable_VR_headsets_ETRA19}. In comparison to VOG and PSOG, EOG-based methods~\cite{shimizu2016eye, bernal2022galea}, as well as SSC-based methods~\cite{whitmire2016eyecontact}, commonly put no requirement on light emitters or receivers. The former approaches eye tracking by placing electrodes around the eyes and then measuring voltage changes during eye movement, whereas the latter realizes gaze estimation by requiring users to wear wired contact lenses and tracking the lenses in a magnetic field. EOG-based methods can be easily implemented in HMDs and can track eyes, even when closed, although with less accuracy~\cite{adhanom2023review}. In contrast, SSC-based methods provide excellent accuracy but are often challenging to deploy. Eye tracking may also be approached with specialized devices and algorithms. One example is achieving gaze estimation using an adaptive optics scanning laser ophthalmoscope, a device that can image the retina at high resolution and high frame rate~\cite{r-slam_iccv_2021}. This approach models eye tracking as a joint estimation of retina motion and appearance, similar to simultaneous localization and mapping (SLAM~\cite{thrun2008simultaneous}), achieving an accuracy of below $\frac{1}{60^{\circ}}$ at 1 kHz sampling rate.  

\subsubsection{Datasets} 
\label{subsubsec:dataset}
While different types of eye-tracking hardware are essential for obtaining eye movement data, researchers often publish publicly available eye-tracking datasets to facilitate the benchmarking of data-driven methods. Data quality, sample size, and context-dependent information are essential for the performance and robustness of the gaze estimators. To this end,~\cite{NVGaze_chi19} proposed two datasets satisfying such criteria, including a synthetic one using anatomically informed eye and face models and a real-world dataset from 35 participants for near-eye gaze estimation, and showed that their trained CNNs perform with accuracy losses of $2^{\circ}$ and $0.5^{\circ}$ for person-independent and personalized setups, respectively. \cite{OpenEDS_Open_Eye_Dataset, garbin_eyetrackingdataset_etra20} introduced the OpenEDS dataset, which was collected from 152 participants in an immersive virtual environment, for both gaze estimation and eye region segmentation. The dataset covers a diverse range of data types, including pixel-level eye region annotations, unlabeled eye images, video sequences, and point clouds. 

In most cases, visual attention is raised in a bottom-up pattern driven by visual stimuli during data collection. \cite{Jiang_2020_CVPR} presented a novel dataset called IQVA (i.e., immersive question-directed visual attention) in which tasks drove visual attention in a top-down style. The dataset consists of eye-tracking data from 975 HMD video clips, each annotated by 14 participants. \cite{360_event_detection_Agtzidis_MM_19} published an eye-tracking dataset for 360$^{\circ}$ videos collected from 13 observers together with an eye event annotation pipeline and a rule-based eye event classifier for event labeling. \cite{fuhl2021TEyeD} introduced the largest unified public eye dataset, gathered using seven different HMDs in the real world, VR, and AR, from 54 participants. The dataset encompasses various types of eye data, including 2D and 3D landmarks, eye segmentation information, and 3D eyeball annotations. Further, \cite{jin2022you} presented a new taxonomy for VR videos that relies on three metrics: camera motion, video quality, and dispersion of region of interest (ROI), along with a dataset on head and gaze behaviors from 100 participants and 27 videos. 

Multimodal datasets also exist, such as the OpenNEEDS~\cite{OpenNEEDS_dataset_etra21}, which includes 44 participants exploring two virtual spaces and contains data from not only the eyes but also the head, hands, and scenes. Furthermore, the VREED dataset for emotion recognition~\cite{tabbaa2022VREED} comprises data from various modalities, including eye movement, electrocardiogram (ECG), and galvanic skin response (GSR), which were recorded from 34 subjects while watching 360$^{\circ}$ videos. Another multimodal dataset, EgoBody, including eye gaze, head motion, and gesture data~\cite{zhang2022egobody}, also emphasizes human body reconstruction from an egocentric view in VR. 

\paragraph{Synthetic Data}
Collecting high-quality eye-tracking data can be costly and time-consuming. This has led to the development of synthetic 3D models such as UT Multi-View~\cite{sugano2014learning}, SynthesEyes~\cite{wood2015rendering}, and UnityEyes~\cite{wood2016learning}, as well as generative algorithms~\cite{Wang_2018_CVPR} and data augmentation approaches~\cite{Yu_2019_CVPR}. While few of these works are specifically tailored for HMDs, some concepts can be adapted for eye data generation in immersive VR. 

\paragraph{Data Annotation}
An essential and costly step during and after data collection is data annotation. \cite{RCEA-360VR_emotion_annotation_in360vids_chi21} studied collecting emotion annotations with eye-tracking data in VR, which is typically annotated retrospectively and can be discrete and time-consuming. By leveraging previous visualization frameworks, HaloLight and DotSize~\cite{xue2020designing}, the authors achieved real-time continuous emotion annotation during data collection and provided temporal labels for downstream tasks. 

\subsubsection{Eye Region Segmentation} 
\label{subsubsec:segmentation}
This section provides an overview of algorithms related to gaze estimation on HMDs, beginning with eye region segmentation. Semantic segmentation of the eye region is crucial for gaze estimation and is often the starting point for many gaze estimation algorithms. Early eye region segmentation algorithms like~\cite{wildes1997iris} are mainly driven by iris texture and sclera extraction~\cite{domain_adaptation_for_eye_segmentation_ECCVW20}. Nowadays, there is an increasing shift towards multi-region eye segmentation, which is typically realized through deep learning. Unlike other semantic segmentation tasks, eye region segmentation is often challenging due to low-resolution images and blurs.

The encoder-decoder network called EyeNet~\cite{eyenet_segmentation_ICCVW19} addresses these issues by using residual blocks in both encoder and decoder to improve gradient flow and using convolutional block attention modules (CBAM~\cite{woo2018cbam}) to enhance boundary sharpness and accuracy, accomplishing a total score of 0.97 on the EDS evaluation metric. \cite{Fuhl_2019_ICCV} explored the applicability of CycleGANs~\cite{zhu2017unpaired} for eye segmentation and suggested three different generative adversarial networks (GANs) for segmentation, image refinement, and image generation, respectively, that were trained with cyclic loss to prevent the discriminator from overfitting. Conventionally, eye segmentation is vulnerable to occlusion caused by eyelids and eyelashes. \cite{ellseg_semantic_seg_for_gaze_tracking_TVCG21} addressed this challenge with a segmentation framework, EllSeg, that is robust against occlusions, which can be implemented jointly with other pupil and iris ellipse segmentation methods in an encoder-decoder pattern. 

\paragraph{Lightweight Eye Segmentation}
Computational resources, such as computational power, memory, and storage, are often bottlenecks of HMDs. \cite{Boutros_2019_ICCV} alleviated the computational burden on embedded systems by introducing their multi-scaled segmentation network (Eye-MMS) based on a multi-scale interconnected CNN. This approach achieved a 3\% loss in accuracy compared to the original model while successfully reducing the number of model parameters from 6,574k to 80k. \cite{eyeseg_ECCVW20} proposed a lightweight encoder-decoder segmentation network, EyeSeg, which adopts a customized loss function to tackle labeled data shortage. EyeSeg contains only 190k parameters while yielding a 94.5\% mean intersection over union (mIOU) on the OpenEDS dataset. Another lightweight segmentation network, RITnet~\cite{chaudhary_ritnet_iccvw_19}, combines U-Net~\cite{ronneberger2015u} and DenseNet~\cite{huang2017densely}, which is only 0.98 MB and allows eye segmentation at 300 Hz in real-time, with an mIOU of 95.3\%, on the OpenEDS dataset.

\paragraph{Dealing with Data Shortage}
Another challenge in eye region segmentation is the lack of high-quality eye images. \cite{D-ID-Net_identity_preserving_image_generation_iccvw19} overcame this by generating eye images from semantic segmentation with their novel model called D-ID-NET. The pipeline of this model consists of two phases: in the first phase, a domain network (D-Net) synthesizes identity-irrelevant images from semantic labels, and in the second phase, identity-specific information is introduced into the images by an identity-specific network (ID-Net). Both networks are CNNs with the same structure but different training strategies. \cite{Buehler2019Seg2Eye} implemented a similar concept with their model Seg2Eye, which generates content-preserving eye images from semantic segmentation, resembling style transfer~\cite{gatys2015neural} in the sense that semantic segmentation defines the content of generated images. In contrast, their styles are controlled by style features extracted from images of the target person. 

\cite{domain_adaptation_for_eye_segmentation_ECCVW20} extensively studied domain adaptation for eye segmentation when only a few source images are annotated, while most data are not labeled. They systematically investigated the impact of annotated data by training the model in supervised, unsupervised, and semi-supervised manners, respectively, and varying the amount of labeled eye images in the target domain during training. More recently, an image-to-image translation network, GeoMaskGAN~\cite{lu2022geometry}, was introduced, and it accounts for geometric consistency and consumes a pair of eye images and an eye segmentation mask as input, and outputs a new pair while reducing the geometric gap between translated images and original ones. 

\subsubsection{Gaze Estimation} 
\label{subsubsec:estimation}
Gaze estimation is the core focus of most eye-tracking studies, as this information is essential for various follow-up tasks. In this subsection, we do not differentiate between stationary and HMD eye tracking, as most gaze estimation algorithms are not specifically tailored for HMDs. Figure~\ref{fig:sec4.1_methods} broadly categorizes gaze estimation methods.

\begin{figure}[h]
\usetikzlibrary{trees}
\centering
\scalebox{0.64}{
\begingroup 
\color{black} 
\begin{tikzpicture}[
  man/.style={rectangle,draw,fill=blue!20},
  woman/.style={rectangle,draw,fill=red!20,rounded corners=.8ex},
  nobackground/.style={},
  hardware/.style={rectangle, draw, fill=blue!20, align=center},
  algorithm/.style={rectangle, draw, fill=red!20, align=center},
  grandchild/.style={grow=down,xshift=1em,anchor=north,
    edge from parent path={(\tikzparentnode.south) -- (\tikzchildnode.north)}},
  grandgrandchild/.style={grow=down,xshift=1em,anchor=north,
    edge from parent path={(\tikzparentnode.south) -- (\tikzchildnode.north)}},
  zero/.style   ={level distance=15ex},
  first/.style  ={level distance=15ex},
  second/.style ={level distance=15ex},
  level 1/.style={sibling distance=10em}]
    \coordinate
      node[nobackground]{\textbf{Gaze Estimation Methods}}
    [edge from parent fork down]
    child[zero]{node[hardware]{\begin{tabular}{c} VOG: video-\\oculography \end{tabular}}
      [edge from parent fork down]
      child[first, right=6em]{node[algorithm]{model-based}
        [edge from parent fork down]
        child[second, right=1em]{node[algorithm]{{\begin{tabular}{c} shape-based \\ \hline \cite{ahuja_eyespyVR_lowcost_IMWUT18, feng2022real, lu2022neural, beyond_10khz_paper_TVCG2021} \end{tabular}}}}
        child[second, right=4em]{node[algorithm]{{\begin{tabular}{c} corneal reflection-based \\ \hline \cite{Stengel2015cheap, low_cost_gaze_tracking_google_cardboard_vrst16, EyeMR_lowcost_prototyping_Cardboard_ETRA18} \end{tabular}}}}
        }
      child[first, right=5em]{node[algorithm]{{\begin{tabular}{c} appearance-based \\ \hline \cite{Ranjan_2018_CVPR_Workshops, Yu_2020_CVPR, benefit_of_temporal_info_for_appearace_based_gaze_est_etra20, Cheng_2018_ECCV, stojanov2022benefits} \end{tabular}}}}
      child[first, right=10em]{node[algorithm]{{\begin{tabular}{c} hybrid \\ \hline \cite{park2018deep, park2018learning} \end{tabular}}}}}
    child[zero]{node[hardware]{\begin{tabular}{c} PSOG: photosensor-\\oculography \\ \hline \cite{optical_gaze_tracking_ismar20, power_efficient_ET_sensor_for_portable_VR_headsets_ETRA19} \end{tabular}}}
    child[zero]{node[hardware]{\begin{tabular}{c} EOG: electro-\\oculography \\ \hline \cite{shimizu2016eye, bernal2022galea} \end{tabular}}}
    child[zero]{node[hardware]{\begin{tabular}{c} SSC: scleral\\search coil \\ \hline \cite{whitmire2016eyecontact} \end{tabular}}};
\end{tikzpicture}
\endgroup 
}
\caption{\label{fig:sec4.1_methods} A general taxonomy of gaze estimation methods from Section~\ref{sec_4_1}.}
\end{figure}

\paragraph{Model-based Gaze Estimation}
Gaze estimation algorithms can be broadly classified into two categories: model- and appearance-based~\cite{hansen2009eye}. Model-based methods are sometimes also referred to as feature-based. Model-based techniques rely on local geometric eye features such as contours and eyeball models, and they often provide high accuracy, while a common disadvantage is the limited range of operation~\cite{zhang2019evaluation}, which is presumably not a central concern for HMDs since sensors are close enough to the eyes. Model-based algorithms can be further categorized into corneal reflection-based (glint-based) and shape-based (glint-free), depending on whether additional light sources are needed or not~\cite{zhang2015appearance}. Glint-based methods often compute the location of the cornea center using the Purkinje reflection of infrared light~\cite{hennessey2006single}, while a typical workflow of glint-free methods~\cite{wood2014eyetab} often begins with eye landmark detection, and then geometric features are fitted to serve gaze estimation in the downstream. 

Previously, gaze estimation methods were often heavyweight and therefore could not function at a high rate with eye tracking. To address this, \cite{feng2022real} introduced a novel model-based algorithm based on event-driven eye segmentation. The model tracks events (i.e., changes in pixel brightness level) to predict ROIs of near-eye images, whose resolutions reduce to 18-32\% compared to original images. Then, eye region segmentation is carried out on the reduced eye images, and gaze estimation is realized upon segmentation. The model can operate at 30 Hz on a mobile device at an accuracy of 0.1$^{\circ}$-0.5$^{\circ}$. \cite{beyond_10khz_paper_TVCG2021} also used motion events to promote the frame rate of gaze estimation by placing event cameras close to the eyes. While conventional model-based pupil detection algorithms approach basic pupil tracking, the motion events are utilized to update pupil location at high frequency. On an event-based eye dataset, the system achieves an accuracy of 0.45$^{\circ}$-1.75$^{\circ}$ at a frame rate of over 10 kHz. Another deficiency of most model-based algorithms is that they do not consider pupil location in 3D, which can negatively influence eye-tracking accuracy. \cite{lu2022neural} proposed a solution with a 3D pupil localization model that utilizes an advanced anatomical eyeball model and accounts for the error caused by corneal reflection. The authors achieved 47.5\% and 18.7\% error reductions for 3D pupil localization and gaze estimation, respectively, compared to prior works. 

\paragraph{Appearance-based Gaze Estimation}
A recent shift towards appearance-based methods has been observed. In contrast to model-based techniques, appearance-based methods are conceptually and structurally simpler: they typically use machine learning to directly learn gaze direction from photographic eye appearance (images). Prior appearance-based methods~\cite{mora2013person, krafka2016eye, wang2016appearance} generally relied on basic machine learning, such as linear regression, support vector machine, and random forest. Nowadays, appearance-based methods often utilize deep learning and CNNs. 

One example of advanced appearance-based models improves the robustness against head pose by head pose clustering~\cite{Ranjan_2018_CVPR_Workshops}, and in contrast to former works, where a dedicated network is needed for each head pose, there is a branched structure where most layers are shared while only a few final dense layers are specified for each head pose. To address the data shortage,~\cite{Yu_2020_CVPR} suggested an unsupervised representation for gaze estimation for unlabelled eye images, and calibration works in a few-shot manner. The authors jointly trained two networks: one for learning gaze representation and another for gaze redirection, utilizing warping field regularization to prevent overfitting and distortion. Unlike prior works that learn embeddings in high-dimensional space, this model captures 2D embeddings that are linked to clear physical meaning, namely, eyeball yaw and pitch. With solely 100 calibration samples, the model could reach an accuracy of 7$^{\circ}$-8$^{\circ}$. 

Most appearance-based gaze estimation methods only utilize static eye images and often overlook temporal traces of gaze, which contain important information. \cite{benefit_of_temporal_info_for_appearace_based_gaze_est_etra20} demonstrated the benefit of temporal gaze sequence by implementing a many-to-one CNN-recurrent model for gaze estimation. Incorporating temporal information, the researchers achieved an average error reduction of up to 19.78\%, specifically 16.91\% for the horizontal axis and 23\% for the vertical axis. In another appearance-based model,~\cite{Cheng_2018_ECCV} introduced a new asymmetric regression-evaluation network (ARE-Net) that exploits the asymmetry of human eyes. The model consists of an asymmetric regression network (AR-Net) that estimates 3D gaze direction and evaluation networks (E-Net), which assess the performance of each eye and adjust the regression strategy accordingly. While a model considering a single eye can achieve an accuracy of 6.3$^{\circ}$ on the MPIIGaze~\cite{Zhang2019MPIIGaze} dataset, the proposed method decreased the error to 5.0$^{\circ}$. Moving to the context of eye tracking in immersive VR, a common problem is HMD slippage, which can degrade the gaze estimation accuracy of appearance-based models due to the camera placement sensitivity. To mitigate this issue, \cite{stojanov2022benefits} incorporated depth information parallel to eye appearance, using two approaches to improve robustness against fitting and slippage. While the first approach combines features from the eye image and depth map channels, the second approach utilizes the so-called transformer-based cross-modal attention block~\cite{zhang2022can}, which significantly enhances the model's generalizability. 

\paragraph{Hybrid Gaze Estimation}
Besides appearance-based and model-based techniques, gaze estimation techniques that combine both have also drawn attention, often referred to as hybrid models. These methods~\cite{park2018deep, park2018learning} commonly use deep learning models to extract eye geometric features and then map the features to gaze positions. 

\paragraph{Calibration}
A procedure closely related to gaze estimation is calibration, which is a prerequisite for accurate eye tracking. However, when deployed in HMDs, calibration can be disrupted by head and body movements. Therefore, recalibration is often necessary during continuous use. Prior work focused on these aspects, for instance, in terms of providing a more immersive and consistent user experience, \cite{gaze_beh_interacted_obj_VR_ETRA19} analyzed the timing and probability of fixations by exploiting the correlation between gaze location, virtual objects, and hand interaction. Based on the gaze patterns, an implicit and consecutive recalibration is necessary during HMD use.  

\subsubsection{Eye Event Classification} 
\label{subsubsec:event}
For data visualization and further offline analysis, eye movements are often divided into a sequence of consecutive events, like fixations, saccades, and smooth pursuits. This procedure is often referred to as eye movement event classification, segmentation, or event detection. Fixations are stable eye movements that occur when the eyes focus on a specific area or volume, whereas saccades are rapid and ballistic movements that switch between consecutive fixations. These two eye movements are combined in the temporal dimension to create visual scanpaths. In contrast to fixations and saccades, there are other, more fine-grained eye movements, such as smooth pursuits, which are defined as eye movements that follow a moving target. 

Traditional classification methods~\cite{salvucci2000identifying}, which were designed for 2D monitor-based scenarios, cannot be directly applied to eye movement event classification in VR HMDs due to the change in frame of reference~\cite{EM_classification_HMD_360_agtzidis_and_dorr_ETRA19}. The authors suggested two approaches to alleviate this problem: the first approach aims to transplant 2D monitor-based classification algorithms into 3D Cartesian space. At the same time, the second approach projects HMD data onto a 2D space, allowing traditional methods to be used without modification. 

Like other gaze-related tasks, eye event classification in VR suffers from data vacancy. To address this, prior work explicitly defined primary and secondary eye events and then proposed a two-stage annotation pipeline~\cite{360_event_detection_Agtzidis_MM_19}, and developed a rule-based eye event classifier for automatic event labeling. Furthermore, \cite{rolff2022saccades} introduced a classification strategy that redefines event detection as a time-to-event problem, and the newly proposed strategy can estimate the remaining time until the next saccade and predict eye behavior in the near future. 

\subsection{Eye-based Human-Computer Interaction in VR} 
\label{sec_4_2}
\begin{figure}[h]
  \centering
  \includegraphics[width = \linewidth]{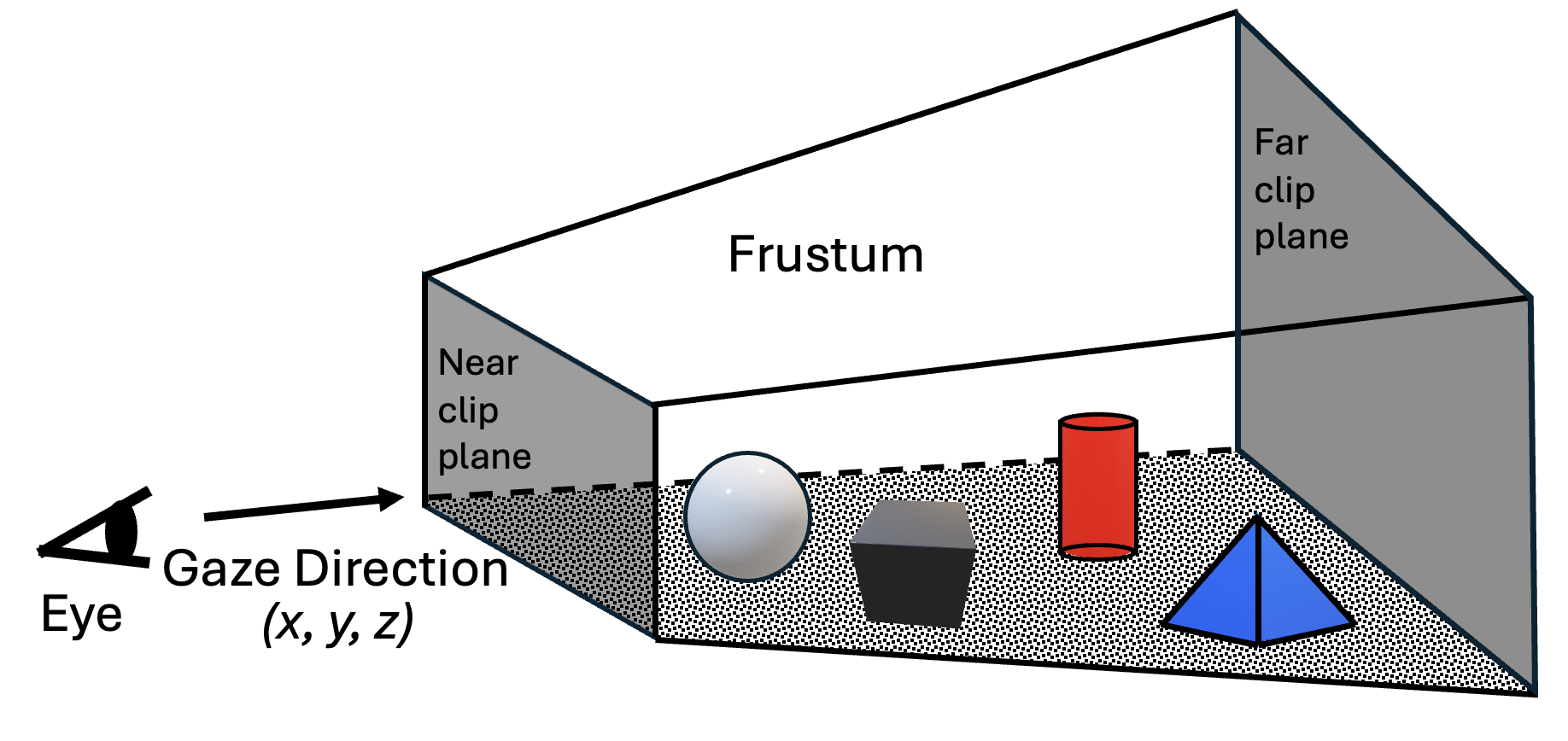}
  \caption{Representation of a general setup for eye tracking in VR.}
  \label{et_vr_sketch}
\end{figure}

The advancements in hardware, datasets, and algorithms have opened up numerous opportunities to enhance HCI in immersive virtual environments. In this section, we examine research in eye-tracking-based HCI within the context of VR. We begin with exploring recent works related to eye-based interaction, explicitly addressing the challenges and potential solutions in Section~\ref{subsubsec:vr_interaction}. Subsequently, we delve into predictive gaze analyses in Section~\ref{subsubsec:predictive_gaze_analysis} by encompassing prediction and visualization techniques. Lastly, Section~\ref{subsubsec:rendering_techniques} provides an overview of eye-based rendering techniques that enhance the real-time performance of VR devices and improve the user experience. Figure~\ref{et_vr_sketch} visualizes the general setup for eye tracking in VR, especially for HMDs. 

\subsubsection{Interaction in VR}
\label{subsubsec:vr_interaction}
Inherent characteristics of virtual environments provide users with novel opportunities for interaction. Many interaction methods require additional controllers, such as handheld ones. In contrast, gaze-based interaction can be executed solely by using built-in or integrated eye trackers in a hands-free manner. However, numerous challenges exist concerning precision, time efficiency, and simplicity. In this section, we focus on works that address existing challenges around eye-based interaction techniques in VR. Table~\ref{tab:sec_4_2_1} provides an overview of papers in this section. 

\begin{table}[!ht]
\scriptsize
\centering
\caption{A broad overview of papers in Section~\ref{subsubsec:vr_interaction}.}
\label{tab:sec_4_2_1}
\begin{tabular}{ll}
  Contribution & Characteristics \\ \hline \hline
    \multirow{8}{*}{\begin{tabular}[c]{@{}l@{}} \\User interface \\ design\end{tabular}} &
\cite{reiter2022look} Hand-attached menu interacted with wrist and gaze \\ 
   &
\cite{choi2022kuiper} Menu placement on Kuiper Belt region \\ 
   &
\cite{kim2022lattice} Menu design incorporating lattices as a guide \\ 
     &
\cite{sticky_pie_scale_inv_marking_menu_chi21} StickyPie: Solution for overshooting and false activation \\
   &
\cite{eyegaze_wheelchair_control_vr_etra20} Wheelchair control with waypoint navigation  \\ 
  &
\cite{yi2022gazedock} GazeDock: View-fixed menu with personalized selection \\ 
  &
\cite{lee2022vrdoc} VRDoc: Reading interface design \\ 
  &
\cite{10.1145/3494968} SwiVR-Car-Seat: Car interface design \\ 
  &
\cite{bionictracking_using_gaze_VR_ECCVW20} Bionic Tracking : Cell tracking in 3D volumetric data 
\\ \hline
  \multirow{14}{*}{\begin{tabular}[c]{@{}l@{}} \\ \\ Target selection\\ \& manipulation \end{tabular}} &
\cite{gaze-asst-selection-sidenmark-chi20} Outline Pursuits: Gaze- and object-based selection \\  
   &
\cite{Radi-Eye_chi21} Radi-Eye: Discrete and continuous selection \\ 
   &
\cite{eyeseethrough_interaction_tech_ieeevr19} EyeSeeThrough: Merging confirmation and selection \\ 
   &
\cite{eye_head_gaze_pointing_uist19} Eye\&Head: Head-supported gaze actions \\ 
   &
\cite{gaze-enhanced_menus_in_VR_vrst20} Comparison of gaze in selection and drawing tasks \\ 
   &
\cite{pointingby_gaze_head_foot_in_VR_ETRA19} Comparison of gaze, foot, head, and mouse interaction \\ 
   &
\cite{PinchClickDwell_etra21} Interaction with dwell time and pinch movement \\ 
   &
\cite{mutasim2022performance} Saccade-based confirmation method  \\ 
 & 
\cite{sidenmark2022weighted} Weighted Pointer: Error-aware gaze-based interaction \\ 
 & 
\cite{9583829} Fatigue analysis for gaze- and controller-based selection
\\ 
 & 
\cite{wei2022label} Gaze-based label guidance for object locating task
\\ 
&
\cite{Pfeuffer_etal_2017} Gaze+Pinch: Gaze for selection with gestures
\\
&
\cite{liu_etal_sui_2020} Eye-gaze-based object rotation and manipulation 
\\
 & 
\cite{meng2022exploration} Comparison of hand-free text selection task
\\ \hline
  \multirow{4}{*}{Text entry} &
\cite{he2022tapgazer} TapGazer: Finger tapping and eye gaze \\ 
   &
\cite{blinktype_text_entry_VR_ismar20} Interaction methods using blink or neck motions 
\\  
   &
\cite{BCI_gaze_textextry_VR_iui18} SSVEP: Brain-computer interfaces + eye tracking \\ 
   &
\cite{gaze_typing_in_VR_impact_of_many_factors_ETRA18} Dwell\&click-based interaction \\ \hline
  \multirow{4}{*}{\begin{tabular}[c]{@{}l@{}}Disambiguation\\ \&\\ Depth\end{tabular}} &
\cite{monocular_gazedepth_using_VOR_ETRA19} VOR-gain-based depth estimation \\ 
   &
\cite{Predicting_gaze_dpth_good_for_gazecontingent_ETRA18} Regression model with vergence-based features \\ 
   &
\cite{3dgaze_interaction_through_Vor_depth_est_chi19} Depth estimation with VOR gain \\ 
   &
  \cite{headgaze_interaction_ieeevr20} Comparison of disambiguation techniques \\ \hline 
  \multirow{3}{*}{Intent} &
\cite{predicting_task_basedon_eye_movements_VR_etra20} Predict tasks based on eye movements \\ 
   &
\cite{David-John_intent_VR_etra21} Logistic regression with focal attention and eye gaze \\ 
   &
\cite{lost_in_style_chi19} LSTM model based on sequence of eye gaze \\ \hline
  \multirow{5}{*}{\begin{tabular}[c]{@{}l@{}}Hand \\ redirection\end{tabular}} &
\cite{sparse_haptic_proxy_chi17} Sparse Haptic Proxy: Mimic haptic feedback \\ 
   &
\cite{reachplus_haptics_with_gaze_VR_UIST20} REACH+: Physical interaction in VR to enhance realism   \\
   &
\cite{blink-suppressed_hand-redirection_ieeevr21} Hand redirection exploiting blink motion of the eye\\ 
   &
\cite{gaze_supported_3d_obj_manip_VR_chi21} Object manipulation tasks beyond arm reach \\ 
   &
\cite{sendhilnathan2022detecting} Error detection in gesture input with the help of gaze \\ \hline
  \multirow{2}{*}{\begin{tabular}[c]{@{}l@{}} \\Walking \\ redirection\end{tabular}} &
\cite{blinks_and_redirected_walking_algos_vrst18} Discrete rotation technique for both blind and open eyes \\ 
   &
\cite{stein2022eye} Prediction for users' positions with LSTMs \\ 
   &
\cite{9644362} Prediction for users' short- and long-term positions \\ \hline
\end{tabular}
\end{table}

\paragraph{User Interface Design}
\label{par:user_interface_design}
We first present a review of the design of user interfaces based on eye tracking in VR. The primary goal of these interfaces is to enable robust interaction while minimizing the time and effort required. Several works have addressed this issue. For instance, \cite{reiter2022look} introduced a hand-attached menu design that can be navigated with wrist rotation and interacted with the eyes, freeing up one hand for other tasks. \cite{choi2022kuiper} proposed a solution to the ``Midas Touch'' problem~\cite{jacob1990you}, a phenomenon that occurs when users' gaze falsely interacts with an object or menu item during a search task, by defining the gazing region between $25^{\circ}$ and $45^{\circ}$ as the ``Kuiper Belt'' which is not frequently targeted by the users during interaction due to being outside of the comfortable eye movement region~\cite{DGaze_gaze_prediction_tvcg20}, but still within the physically reachable area~\cite{stahl2001eye,stahl1999amplitude}. By placing menu items within the Kuiper Belt, the number of false inputs during the search task was significantly reduced, avoiding the Midas Touch. 

Furthermore, a different menu design that aims to improve the precision of gaze-based selection and reduce the time required for selection was suggested~\cite{kim2022lattice}. Incorporating lattices as a guiding mechanism for gaze gestures in the menu structure yielded fewer selection errors and shorter selection times than conventional gaze interactive menu designs~\cite{urbina2010pies}. \cite{sticky_pie_scale_inv_marking_menu_chi21} introduced a novel gaze-based VR/AR marking menu named ``StickyPie'' that addressed the limitations in existing eye-tracking interaction techniques, including overshooting and false activation. StickyPie, which provides a scale-invariant marking menu, achieved over 10\% improvement in efficiency compared to RegularPie, a more conventional scale-variant menu design. \cite{wei2022label} presented an alternative interface design for the object localization process with gaze in several VR applications featuring numerous objects, resulting in a reduced task load and time compared to traditional methods. 

Recently, \cite{yi2022gazedock} presented GazeDock as a view-fixed peripheral menu that is automatically displayed when the user's gaze moves to the menu region. GazeDock, with a personalization and optimized selection algorithm, achieved an average selection time of 471 ms and a false trigger rate of 3.6\% while being preferred over dwell- and pursuit-based approaches. \cite{lee2022vrdoc} proposed a virtual interface, VRDoc, for a reading task that might be useful for office workers. VRDoc incorporates eye-based interaction to reduce document selection and navigation times and required effort compared to existing VR reading interfaces. \cite{bionictracking_using_gaze_VR_ECCVW20} designed an eye-tracking-based user interface, Bionic Tracking, to facilitate object tracking by exploiting smooth pursuits in VR and achieved a speed-up of 2-10 times compared to traditionally used 2D point-and-click methods. 

It is also essential to design user interfaces and interaction modalities that consider the specific needs, domains, and setups. To this end, SwiVR-Car-Seat~\cite{10.1145/3494968} explores the impact of vehicle motion on VR interaction in automobiles with an experiment using a low-cost rotating seat to observe the effect of vehicle rotation on touch, gesture, gaze, or speech interactions. The results indicate that vehicle motion has a detrimental impact on gaze and gesture interactions, whereas touch and speech interactions were more resilient. Furthermore, \cite{eyegaze_wheelchair_control_vr_etra20} proposed gaze-based control interfaces, including overlay control and continuous control interfaces, and waypoint navigation. In a wheelchair control setup, the semi-autonomous waypoint gaze interface yielded the fastest task completion time for each trial, indicating a more favorable user experience compared to other interaction methods. 

\paragraph{Gaze-based Target Selection and Manipulation}
\label{par:eye_based_selection}
Gaze-based user interface design requires efficient object and region selection to enable a smooth interaction experience. To this end, Eye\&Head~\cite{eye_head_gaze_pointing_uist19} is an approach that separately evaluates only gaze- and head-supported gaze actions for hands-free target selection. It was found that eye-tracking data, combined with head motion, provides greater freedom to users during the selection process, as users' intent is more evident in this setup. In another work, \cite{gaze-asst-selection-sidenmark-chi20} proposed Outline Pursuits, which utilizes gaze to solve object selection problems in occluded virtual scenes. The authors assigned different motions to candidate objects and analyzed the correlation between gaze and object movements. Using Outline Pursuits, the selection process requires less effort and shorter time than traditional ray-casting methods, also providing slightly better accuracy in highly occluded environments. 

Another selection technique, Radi-Eye~\cite{Radi-Eye_chi21}, enables hand-free interaction in virtual smart home applications. Like the previous work, Radi-Eye uses gaze data to perform discrete or continuous target and object selections, while head movement is primarily used to confirm or modify the choice. As Radi-Eye provided a more precise and time-efficient interaction compared to previous works, it also offers essential design insights for immersive VR. Furthermore, \cite{qian_and_teather_2017} evaluated different eye-based selection methods in VR, including eye-only, eye and head combined, and head-only input methods when users were seated, and selection targets appeared in front of the participants, and found that the head-only selection technique was preferable during the interaction, indicating that users' context plays a vital role in interactions in VR.

Gaze+Pinch~\cite{Pfeuffer_etal_2017} is another interaction method in VR, where eye movements are used for target selection and freehand gestures to manipulate them. Without relying on extra controllers, users perceived the proposed method as intuitive. \cite{liu_etal_sui_2020} focused on virtual object manipulations, particularly on rotations using gaze input, and stated that ways of interaction purely by eye gaze can be advantageous for individuals with limited use of hands and arms. \cite{eyeseethrough_interaction_tech_ieeevr19} presented EyeSeeThrough, a novel eye-based interaction technique that integrates both confirmation and selection processes through the line-of-sight direction, and EyeSeeThrough surpasses the performance of two-stage selection methods in terms of time and comfort. Similarly, \cite{gaze-enhanced_menus_in_VR_vrst20} evaluated the effectiveness of gaze-based interaction in virtual handheld menus. The standard selection methods, including dwell time, gaze button, and cursor, were integrated with eye-tracking data and compared to a pointer-based selection method. Evaluations indicated that gaze-based selection reduces physical effort compared to traditional selection methods. Furthermore, \cite{pointingby_gaze_head_foot_in_VR_ETRA19} evaluated the performance of gaze, foot, head, and mouse pointing methods in selection tasks, and found that head input was superior to gaze regarding ease of calibration, effective target width, and throughput, while gaze input performed similarly to foot input. 

Another study examining dwell and pinch movements as alternative versions of click-in devices supporting visual attention-based interaction indicated that pinch gestures could be a viable alternative to conventional button click-based selection methods~\cite{PinchClickDwell_etra21}. Furthermore, \cite{mutasim2022performance} evaluated the performance of a saccade-based selection and confirmation procedure compared to traditional approaches such as dwell and button press, and found that the saccade-based selection was time-efficient, yet error-prone. 

Recently, \cite{sidenmark2022weighted} introduced a weighted pointer, an error-aware gaze-based interaction technique, which is designed to maintain stability in the presence of eye-tracker errors by integrating fallback modalities. Evaluations demonstrated that a weighted pointer, allowing for automatic switching of modalities, is more effective and favorable than techniques that require manual switching. In another recent work, \cite{meng2022exploration} investigated the efficiency of hands-free text selection techniques, including selection using the dwelling, blink, and voice. The target selection using blinks outperformed the other mechanisms regarding time, accuracy, effort, and preference. 

To understand the practical aspects of gaze-based selection, \cite{9583829} analyzed the occurrence of fatigue, comparing selection techniques based on eye gaze and controller input. The findings revealed that prolonged use of eye gaze could lead to fatigue, making it unsuitable for interactions for extended periods. Similarly, \cite{Hou_and_Chen_2021} compared eye- and controller-based selection in VR and indicated that while eye-based interaction has potential, it still has some way to go before becoming a mainstream interaction method in VR due to the stability and precision issues with eye trackers. In another perspective, \cite{Prithul_etal_2022} evaluated hands-free alternatives of controller-based teleportation, and found that teleport activations by winking provided the most viable alternative compared to mouth gestures and dwelling, as it led to fewer selection errors. 

\paragraph{Gaze-based Text Entry}
\label{par:eye_based_text_entry}
Similar to eye-based target-selection techniques, another form of interaction is text input. However, the nature and requirements of text-entry tasks differ and should be considered in the context of VR interactions. Moreover, gaze information can support not only interaction itself but also reveal potential errors~\cite{Peacock_etal_2022_etra}.

To this end, \cite{he2022tapgazer} proposed TapGazer, utilizing finger tapping to type on a virtual keyboard by incorporating gaze assistance for word selection. \cite{blinktype_text_entry_VR_ismar20} presented BlinkType and NeckType text entry techniques, utilizing blinks and neck movements in character selection, respectively, and found that using blinks is more favorable and reaches the highest word-per-minute rate compared to NeckType and dwell-based text-entry techniques. Another approach~\cite{BCI_gaze_textextry_VR_iui18} combined brain-computer interfaces and eye tracking, enabling users to compose ten words per minute and achieving an information transfer rate of 270 bits per second. \cite{gaze_typing_in_VR_impact_of_many_factors_ETRA18} conducted a comparative investigation on applying dwell-time and click inputs in conjunction with gaze as a confirmation mechanism, also analyzing various keyboard visualization options, and found that the most convenient combination was click actions with the entire keyboard. Additionally, \cite{eyeMRTK_toolkit_gaze_interactive_apps_inVR_ETRA19} presented EyeMRTK, an eye-tracking-based toolkit that supports text entry in Unity. 

\paragraph{Gaze-based Disambiguation}
\label{par:user_disambiguation}
Gaze-based interaction methods may suffer from inaccuracy and stability issues due to the imprecise detection of target objects, which can be influenced by factors such as inaccurate eye tracking or the complexity of the virtual environment. We discuss eye-gaze-based disambiguation techniques that address these challenges. 

Interaction methods that rely solely on the intersection of the gaze ray may cause less accurate detection of targeted objects, especially in scenes with objects at multiple depth levels. To address this issue, depth estimation methods based on the eye have been proposed to improve the interaction quality. Most approaches rely on the relationship between target depth and the vestibulo-ocular reflex (VOR) to fix gaze by moving the eye in the opposite direction of the head. \cite{monocular_gazedepth_using_VOR_ETRA19} proposed a VOR-based depth estimation approach using VOR gain obtained by observing pupil centers, as a measure of in-depth computation, rather than the more conventional vergence measure. The VOR gain-based model, which only requires one eye, revealed that its performance is comparable to that of the vergence measure conventionally used in in-depth estimation. \cite{3dgaze_interaction_through_Vor_depth_est_chi19} introduced a VOR-based technique to overcome disambiguation problems during gaze interaction by jointly utilizing eye and head movements to estimate gaze depth, leading to a better performance compared to conventional vergence-based methods in gaze depth estimation for the virtual objects located deep in the scene. 

In another work, \cite{Predicting_gaze_dpth_good_for_gazecontingent_ETRA18} developed a regression model to enhance gaze depth estimation by integrating vergence measures with other features obtained through conventional ray-tracing techniques based on the users' point of regard. The model yielded remarkable accuracy regarding the average deviation from the benchmark depth over a six-meter range, while vergence measures offered accurate prediction only up to a meter. Similarly, disambiguation is also a significant challenge in hands-free interaction. \cite{headgaze_interaction_ieeevr20} evaluated disambiguation techniques such as head gaze, speech, and foot tap in different timing scenarios and found that head gaze outperformed other methods in eliminating disambiguation. 

User intent information is also helpful in addressing disambiguation issues that arise in VR interaction based on eye tracking. By predicting user intent, it is possible to identify forthcoming movements of users and objects with which users will interact, and several works have utilized eye tracking for this. For example, \cite{predicting_task_basedon_eye_movements_VR_etra20} demonstrated the capability of point-of-regard regions to predict four different alignment tasks with cubes, achieving an $F_{1}$-score of $0.51$. \cite{David-John_intent_VR_etra21} employed eye-tracking features to train a logistic regression model and revealed that characteristics of focal attention, gaze velocity, and scanpath dynamics are highly relevant for predicting user intention. Furthermore, \cite{9664291} proposed EHTask, a novel task classification method that utilizes eye and head movement features, and demonstrated that eye-tracking features differ based on the tasks in VR. Additionally, EHTask outperformed state-of-the-art methods in task classification for 2D viewing. Similarly, for intent prediction, \cite{lost_in_style_chi19} introduced an approach that predicts navigation assistance needs based on eye gaze information and long short-term memory networks (LSTMs), achieving a prediction accuracy of over 80\%. 

\paragraph{Gaze-based Hand Redirection}
\label{par:hand_redirection}
Hand redirection methods are essential for addressing the discrepancies between hand movements in virtual and real environments, as perfect alignment between the two is not always possible. Discrepancies can lead to difficulties during hand interactions, such as the inability to interact with the intended objects. Eye-tracking-based methods can help identify the user's intent and direct the hand to the correct location. 

To this end, \cite{sparse_haptic_proxy_chi17} utilized a gaze-based hand redirection method to bridge the gap between real and virtual worlds with their framework, Sparse Haptic Proxy. This framework consists of geometric primitives that mimic the haptic feedback of virtual objects. The authors employed the haptic re-targeting technique~\cite{azmandian2016haptic} that utilizes users' eye and hand behaviors to predict intentions and redirect their hand to a physical proxy, and showed that it achieved an accuracy of 97.5\% for gaze-based user intention prediction, while reporting the maximal acceptable hand redirection angle as $40^{\circ}$. \cite{blink-suppressed_hand-redirection_ieeevr21} proposed Blink-Suppressed Hand Redirection as a hand redirection approach based on a body wrapping algorithm~\cite{sparse_haptic_proxy_chi17}. Unnoticeable instant changes are performed by exploiting the natural blink motions, and the real and virtual hand offset is adjusted with slight modifications when the eyes are open. 

REACH+~\cite{reachplus_haptics_with_gaze_VR_UIST20} combines eye-tracking data and hand motion to determine user intention in redirection frameworks, and aims to overcome challenges through physical interaction in virtual spaces. REACH+ redirects the users' hands to the intended target within arm-reachable range, enhancing the sense of realism. In recent work~\cite{sendhilnathan2022detecting}, error detection in gesture input was accomplished with the help of gaze dynamics, and the gaze patterns following gesture input were analyzed to classify the input into three types: correct input, input recognition errors, and user errors. Utilizing only gaze features such as fixation duration, saccade amplitude, and gaze velocity led to a classification area under the curve of a receiver operating characteristic curve for one-vs-rest score (AUC-ROC-OVR) of 0.78 with a temporal convolutional network (TCN). Considering object manipulation tasks, \cite{gaze_supported_3d_obj_manip_VR_chi21} designed an interaction technique combining hand gestures with gaze information for object manipulation tasks in VR, and evaluated four alternative combination techniques while varying the object distance from the agent in the virtual environment. Combining hand and gaze improved the usability and efficiency of object manipulation tasks in large environments where objects are not within arm's reach. 

\paragraph{Gaze-based Walking Redirection}
\label{par:walking_redirection}
Another challenge in VR interaction, especially during mobile activities, is the limited space in the real world. \cite{blinks_and_redirected_walking_algos_vrst18} proposed a discrete rotation technique to minimize the required area in a real workspace. They first assessed the threshold for the discrete rotation in virtual scenes and obtained $9.1^{\circ}$ and $2.4^{\circ}$ for closed and open eyes, respectively. Then, they applied discrete scene rotation while the agent was walking, reducing the required area by 20\%. An alternative method to facilitate virtual walking activity in constrained physical spaces employed a path-planning algorithm that considered the positions of pre-existing static and dynamic obstacles, including walls and furniture, as well as multiple VR users within a designated room~\cite{infinite_walking_VR_saccadic_redirection_TOG18}. Furthermore, the algorithm utilized saccadic suppression periods to facilitate efficient path planning, and the authors presented a subtle gaze direction technique to increase the number of saccades, thereby considerably enhancing the redirection gain. From a user perception perspective, \cite{brument_etal_sui_2021} studied the influence of translational and rotational motion in the context of rotation gains and found, through gaze analyses, that rotation gains can alter user behavior, which may be important when implementing redirection techniques. 

Predicting future physical positions can help optimize the use of limited physical area~\cite{stein2022eye,9644362}. \cite{stein2022eye} proposed to use future location predictions obtained through an LSTM model utilizing eye tracking, and demonstrated that their model can predict users' position 2.5 seconds in advance, with an average error of 65 cm. A similar approach, relying on the LSTM model and evaluating short-term (i.e., 50 ms) and long-term (i.e., 2.5 seconds) predictions separately, achieved optimal performance in short-term predictions by utilizing position and orientation features~\cite{9644362}. In contrast, adding eye-tracking features resulted in the best performance for long-term predictions. The authors stated that short-term predictions might be valuable for optimizing computational resources for rendering or transmitting bandwidth in streaming activities. 

\subsubsection{Predictive Gaze Analysis}
\label{subsubsec:predictive_gaze_analysis}
Predicting the future gaze locations enhances the efficiency of gaze-based rendering techniques, particularly foveated rendering. Therefore, methods that combine gaze prediction and foveated rendering have gained significant momentum in recent years. This section discusses the algorithms for gaze location prediction, scanpath prediction, and saliency map generation. Table~\ref{tab:sec_4_2_2} provides an overview of the papers in this section. 

\begin{table}[!ht]
\scriptsize
\centering
\caption{A broad overview of papers in Section~\ref{subsubsec:predictive_gaze_analysis}.}
\label{tab:sec_4_2_2}
\begin{tabular}{ll}
  Contribution & Characteristics \\ \hline \hline
  \multirow{6}{*}{Gaze prediction} &
  \cite{DGaze_gaze_prediction_tvcg20} DGaze: CNNs with object, head, and saliency data\\ 
     &
\cite{SGaze_tvcg19} SGaze: Data-driven model for real-time gaze prediction \\ 
   &
  \cite{fixationnet_bulling_tvcg21} FixationNet: Eye\&Head tracking and task-related data \\ 
   &
\cite{gaze_prediction_dyn_str_manip_games_ieeevr16} In-game variables and gaze features  \\ 
   &
\cite{Zhang_2017_CVPR} Deep Future Gaze (DFG): GAN model \\ 
   &
\cite{viewport_forecasting_aivr19} Head orientation prediction with linear SVM \\ \hline
  \multirow{4}{*}{\begin{tabular}[c]{@{}l@{}}Scanpath \\ prediction\end{tabular}} &
\cite{assens2017saltinet} SaltiNet: CNN model with saliency volumes \\ 
   &
\cite{Assens_2018_ECCV_Workshops} PathGAN: GAN model for scanpath prediction \\ 
   &
\cite{martin2022scangan360} ScanGAN360: Scanpath generation from graph \\ 
   &
\cite{zhu2018prediction} Utilization of clustering and graph-based algorithms \\ \hline
  \multirow{6}{*}{\begin{tabular}[c]{@{}l@{}} Saliency map \\ estimation\end{tabular}} &
\cite{John_360Saliency_aivr18} Benchmarking different methods \\ 
   &
\cite{model_based_realtime_vis_3dheatmaps_ETRA16} Multi-perspective 3D saliency map estimation \\ 
   &
\cite{sitzmann_saliency_tvcg18} User-specific behaviors \\ 
   &
\cite{360VideoSaliency_in_HMD} LSTM model with head and saliency features \\ 
   &
\cite{impact_of_360_mov_on_user_att_ieeevr20} User attention in cinematographic VR movies  \\ 
   &
\cite{salient_object_det_360_and_dataset_TVCG20} Salient object detection (SOD) \& 360-SSOD dataset\\ \hline
\end{tabular}
\end{table}

\paragraph{Gaze Prediction}
Foveated rendering methods depend on accurate gaze prediction, which directly impacts the performance and usability of foveated rendering applications. Therefore, researchers have devoted significant attention to exploring and implementing effective gaze prediction techniques, leveraging various methods, ranging from statistical approaches to neural networks. For instance, Deep Future Gaze (DFG)~\cite{Zhang_2017_CVPR} is a GAN-based model that exploits a spatial-temporal CNN as an encoder and anticipates upcoming gaze positions while generating future frames based on the current ones. A by-product of their work is an object search task (OST) dataset, considered one of the most extensive egocentric datasets. \cite{SGaze_tvcg19} proposed SGaze that utilizes eye-gaze information collected by an integrated eye tracker to forecast future gaze locations. SGaze was designed as a data-driven model based on eye-head coordination for real-time gaze prediction, utilizing statistical models that exploit the relationship between gaze and head movements, without requiring any additional hardware or eye-tracker support. This model took into account the latency in eye and head movements, achieving a better angular distance value than the baseline, which used the screen center as the gaze point. More recently, \cite{DGaze_gaze_prediction_tvcg20} presented DGaze as a CNN-based gaze prediction algorithm incorporating dynamic object positions, head movements, and saliency features. Object positions, head velocity data, and previous gaze location sequences are processed using a sequence encoder model to enhance precision. This technique was applied in real-time, and near-future gaze predictions corresponded to short-term predictions, such as those within 200-1000 ms. DGaze achieved better angular distances in dynamic and stationary scenes than SGaze~\cite{SGaze_tvcg19}. 

FixationNet is a neural network utilized to predict near-future gaze locations up to 600 ms in task-oriented virtual experiences, using eye-tracking data, head movements, saliency maps, and task-related information~\cite{fixationnet_bulling_tvcg21}. Evaluations during the visual searching task in VR indicated a strong correlation between fixation positions and head movements, saliency maps, and content-related information. Besides estimating the gaze position, the proposed model outperforms state-of-the-art models in gaze prediction tasks for 150-600 ms in task-oriented and free-viewing experiments. Another gaze prediction method focuses heavily on task-oriented games and exploits the correlation between the game variables in their current states and gaze locations~\cite{gaze_prediction_dyn_str_manip_games_ieeevr16}. Moreover, predicted gaze locations were utilized in the dynamic disparity manipulation method to enhance depth perception in game scenes. \cite{viewport_forecasting_aivr19} predicted the users' future head orientation using past head movements and eye-tracking data with a support vector machine (SVM). This model can predict future viewports without utilizing any information related to the content in spherical videos; hence, it can help improve rendering performance and reduce the transmission load. 

\paragraph{Scanpath Prediction}
Unlike gaze prediction techniques that forecast individual samples, scanpath prediction involves predicting the sequence of gaze points, enabling the estimation of complete human attention to a given scene. While these concepts are related, they serve different purposes and require different models and algorithms for prediction. 

To this end, \cite{assens2017saltinet} proposed SaltiNet, a deep CNN trained first to generate saliency volumes that capture temporal information, and then the scanpath is sampled from the saliency volumes. In another work, \cite{Assens_2018_ECCV_Workshops} designed a novel GAN model, PathGAN, that employs a convolutional-recurrent architecture for both the encoder and decoder to predict scanpath. Recently, \cite{martin2022scangan360} introduced another GAN-based scanpath prediction model, ScanGAN360, which deploys a novel GAN objective function based on dynamic time warping. Unlike the ScanGAN360 \cite{zhu2018prediction}, which suggested generating scanpaths by building a graph from a saliency map, the saliency map is first binarized and clustered into centers. Then, a weighted graph is created with the cluster centers as nodes, and a scanpath can be generated from the graph. 

\begin{figure}[h]
  \centering
  \includegraphics[width=0.7\linewidth]{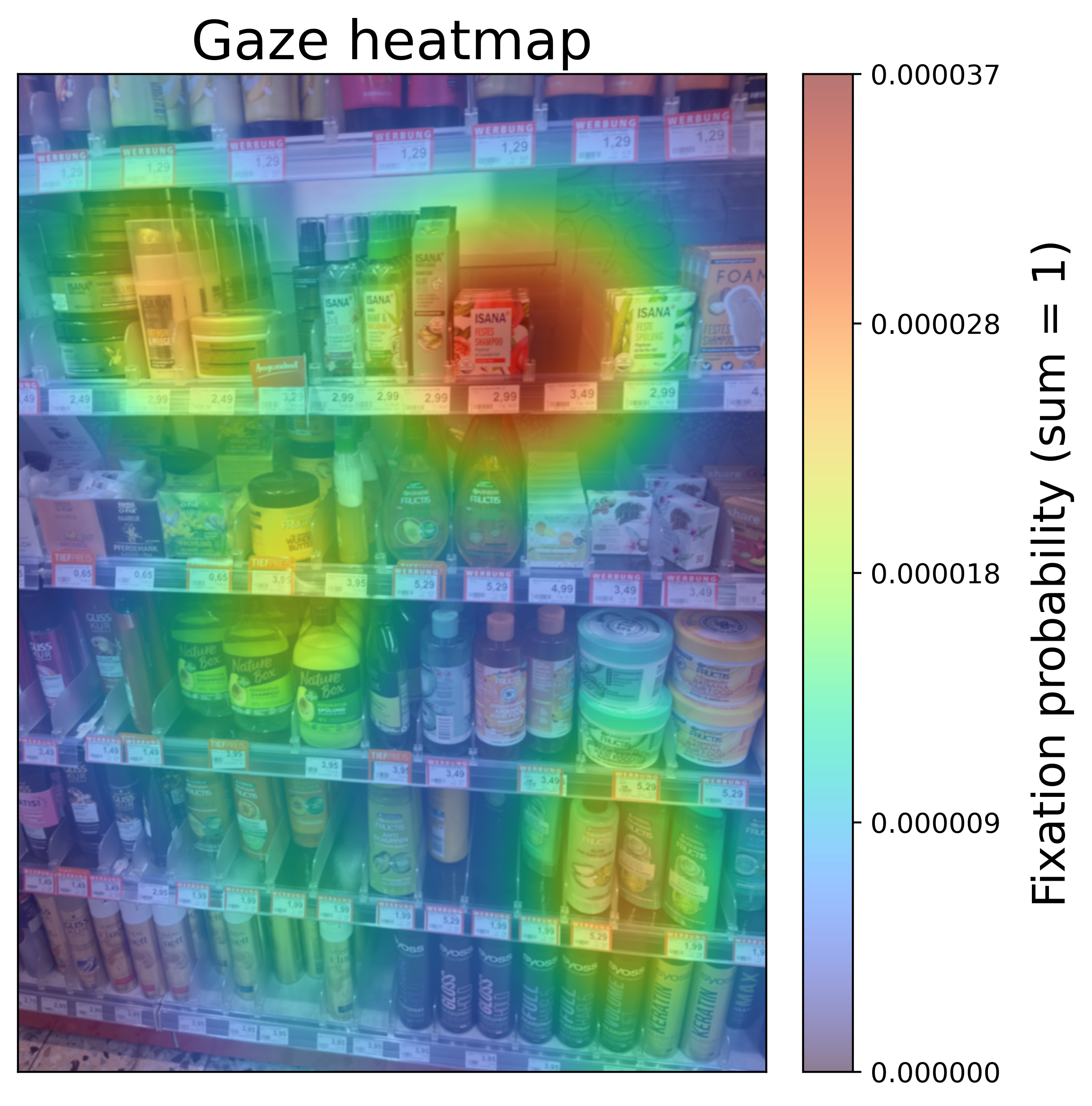}
  \caption{An example of a saliency map, a gaze visualization method that highlights image regions with a color-coding scheme based on visual attention.}
  \label{fig:saliency_example}
\end{figure}

\paragraph{Saliency Map Estimation}
A saliency map, also known as a heat map or attention map, is a visual representation of eye-gaze data on an image or a scene. It serves as a visualization that highlights the regions that attract the most attention from humans. Figure~\ref{fig:saliency_example} illustrates a saliency map, showing the distribution of human attention. Saliency maps lay the foundation for numerous gaze-based VR applications.  

Several approaches exist in the literature for estimating and generating saliency maps. \cite{John_360Saliency_aivr18} benchmarked four different methods for saliency map generation in VR, and suggested utilizing the modified Gaussian kernel~\cite{upenik2017simple} with a scale of 5\% for $360^{\circ}$ saliency map generation due to ease of implementation. \cite{model_based_realtime_vis_3dheatmaps_ETRA16}'s approach addresses challenges such as changing perspectives, dynamically moving objects, and depth of fixations in generating saliency maps for 3D environments. The authors aggregated user gaze data and utilized textures to represent the distribution of visual attention at the object level. Their method offers high-quality saliency maps for multi-perspective eye-tracking recordings. 

\cite{sitzmann_saliency_tvcg18} investigated the difference in saliency patterns between desktop and stereoscopic vision and adapted traditional saliency prediction methods according to user-specific behaviors such as particular fixation biases. \cite{360VideoSaliency_in_HMD} researched predicting head movement using an LSTM model in conjunction with saliency maps. \cite{impact_of_360_mov_on_user_att_ieeevr20} utilized a dataset of 3,259 users watching cinematographic VR movies and measured user attention using saliency maps. Closely related to visual saliency, \cite{salient_object_det_360_and_dataset_TVCG20} aimed at salient object detection (SOD) in $360^{\circ}$ panorama images with eye tracking and proposed a novel dataset, 360-SSOD, including object-level saliency ground truth with balanced semantic distribution compared to existing datasets.

\subsubsection{Rendering Techniques}
\label{subsubsec:rendering_techniques}
The quality of the 3D scene display is paramount for VR devices, as it directly impacts users' overall immersive experiences. Therefore, rendering with high resolution and high frame rates in VR has been very important. However, several challenges exist regarding computational issues for real-time rendering, computational power, and realistic and natural depth perception. To overcome those, different rendering techniques utilizing gaze information have been proposed. In this section, we discuss rendering techniques in three categories based on their objectives: ensuring computational load and power efficiency, optimizing transmission efficiency, and enhancing scene quality. By exploring these categories, we aim to provide a deeper understanding of the current state of the art in rendering in VR and the potential for future developments. Table~\ref{tab:sec_4_2_3} provides an overview of the papers in this section. 

\begin{table}[!ht]
\scriptsize
\centering
\caption{A broad overview of papers in Section~\ref{subsubsec:rendering_techniques}.}
\label{tab:sec_4_2_3}
\begin{tabular}{ll}
  Contribution & Characteristics \\ \hline \hline
  \multirow{6}{*}{\begin{tabular}[c]{@{}l@{}} \\ \\Foveated \\ rendering (FR) \\ \end{tabular}} &
\cite{towards_foveated_rendering_gazetracked_VR} Blur Filter width based on retinal eccentricity \\ 
   &
\cite{eye_domi_guided_fov_rendering_TVCG20} Eye dominance measured with Miles test \\ 
   &
\cite{3DKernel_Foveated_Rendering_for_Light_Fields_TVCG20} 4D-Light areas and gaze-based acceleration \\ 
   &
\cite{Luminance-Contrast-Aware_Foveated_Rendering_TOG19} Content-aware foveated rendering \\ 
   &
\cite{deng2022fov} Neural radiance fields (NeRF) \\ 
   &
\cite{9583738} Foveated depth of field rendering \\ 
   &
\cite{9583734} FR with a subsampling technique for ray tracing
\\ 
   &
\cite{ye2022rectangular} RMFR: Rectangular mapping-based foveated rendering  
\\ \hline 
   \multirow{2}{*}{\begin{tabular}[c]{@{}l@{}}Saccade-based \\ FR \end{tabular}} &
\cite{Arabadzhiyska2017} Polynomial fitting for saccade landing point estimation \\ 
   &
\cite{data_augmentation_for_saccade_landing_point_pred_etra20} Data augmentation for neural networks \\ \hline
  FR assessment &
\cite{is_foveated_perceivable_in_VR_MM17} Foveated rendering assessment \\ \hline
  \multirow{3}{*}{Power efficiency} &
\cite{duinkharjav2022color} Gaze-based approach for colour discrimination \\ 
   &
\cite{refocusable_gp_panoramas_for_IVR_TVCG21} Gaze-based dynamic refocusing for gigapixel panoramas\\ 
   &
\cite{focusVR_IMWUT18} FocusVR: Gaze-based intelligent dimming technique \\ \hline
  \multirow{3}{*}{Streaming} &
\cite{log-rectilinear_transformation_for_fov_360deg_vid_streaming_TVCG21} Log-rectilinear transformation \\ 
   &
\cite{gaze-aware_streaming_nextgen_mobileVR_TVCG18} Codec supporting multi-resolution in a frame \\ 
   &
\cite{chen2022instant} Low-resolution 2D scene processing and 3D mapping \\ \hline
  \multirow{4}{*}{\begin{tabular}[c]{@{}l@{}}Depth sense \\ enhancement \end{tabular}} &
\cite{gaze-contingent_depth_of_field_chi14} Blurring based on focal points and gaze \\ 
   &
\cite{gazecontingent-ocular_parallax_rendering_VR_TOG20} Foveated rendering accounting for ocular parallax   \\ 
   &
\cite{optimizing_depth_perception_VRAR_via_gaze-contingent_rendering_TOG20} Ocular parallax aware gaze-based stereo rendering  \\ 
   &
\cite{gazestereo3d_utilizes_fixation_gaze_est_TOG16} Gradual and stereoscopic depth adjustments \\ \hline 
  \multirow{5}{*} {\begin{tabular}[c]{@{}l@{}} \\ Vergence- \\ accommodation \\ conflict \end{tabular}}     &
\cite{focal_surface_displays_TOG17} Phase-only spatial light modulator (SLM) \\ 
   &
\cite{fast_gaze-contingent_optimal_decompositions_TOG17} Decomposition method with gaze and head motions \\ 
   &
\cite{aizenman2022statistics} Binocular disparities \& screen distance \\ 
   &
\cite{accomodation_invariant-comp_near-eye-display_TOG17} Accommodation-invariant display\\ 
   &
\cite{dunn2019required} Minimum eye-tracking accuracy for varifocal displays\\ 
   &
\cite{batmaz2022effect} Multi-focal and single-focal comparison \\ \hline
\end{tabular}
\end{table}

\begin{figure}[h]
    \includegraphics[width=0.492\linewidth]{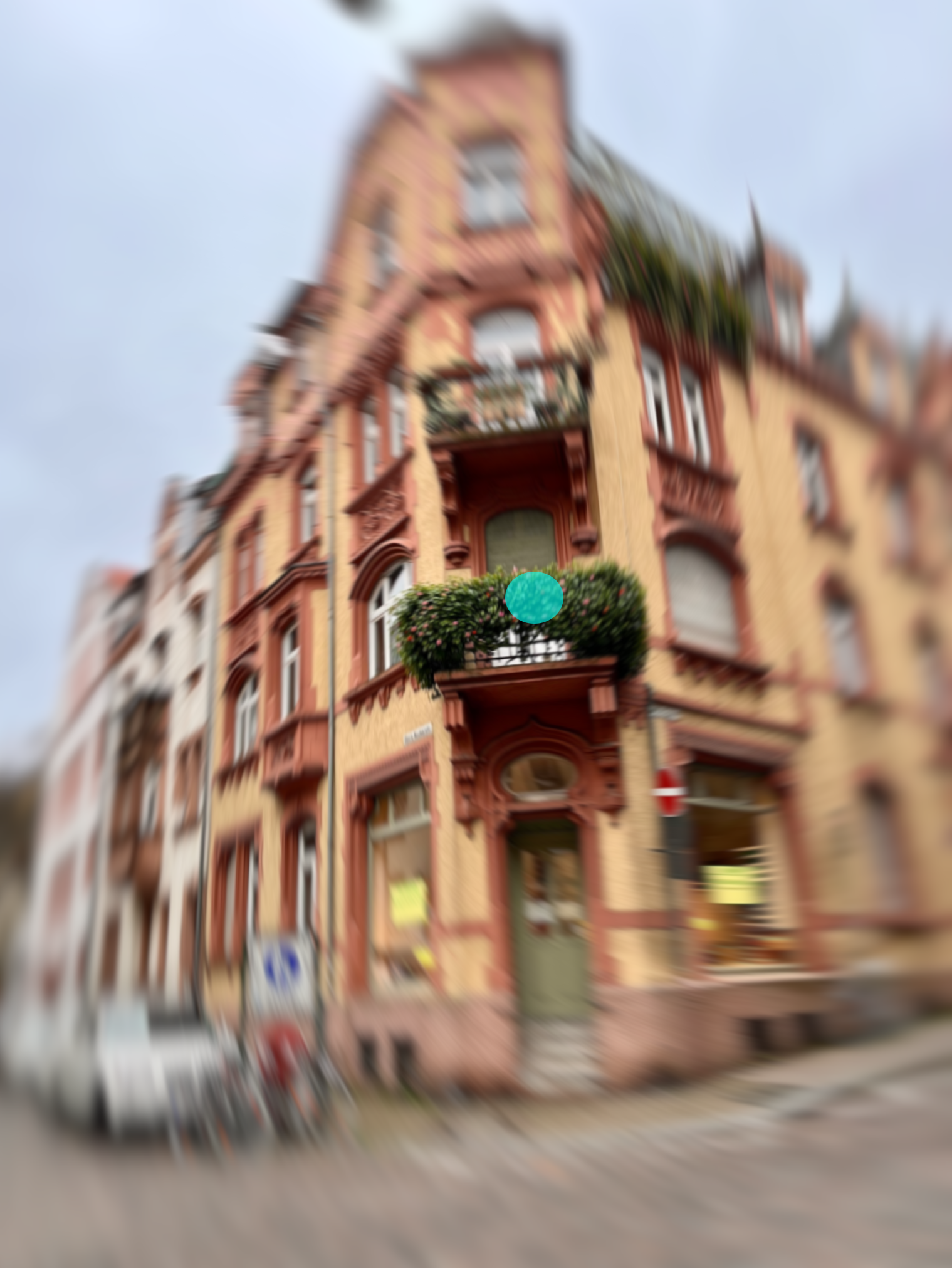}
    \includegraphics[width=0.495\linewidth]{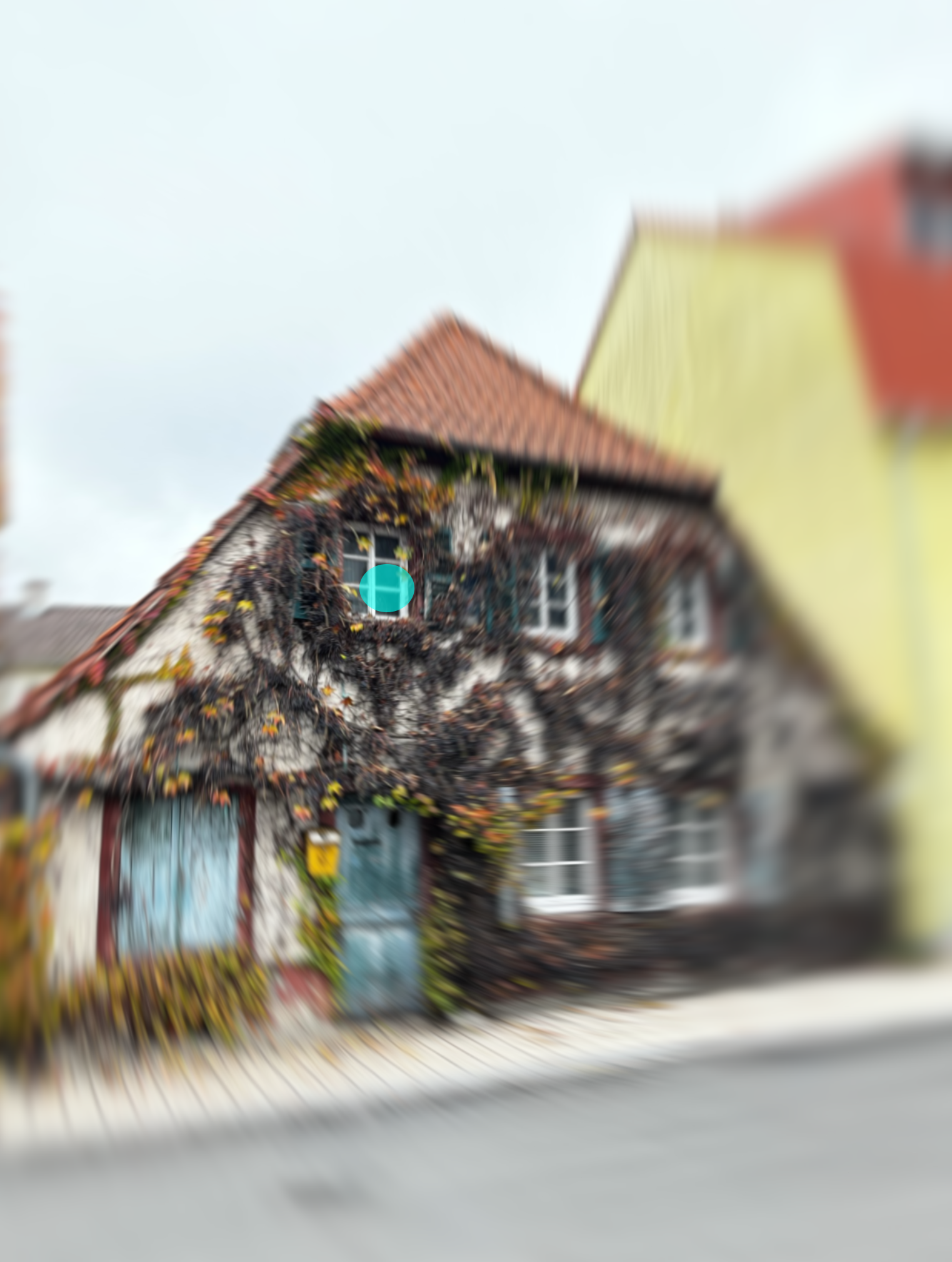}
\caption{Foveated rendering provides a higher level of detail at the viewer's gaze location while reducing the level of detail in the peripheral vision. We depict exemplary gaze locations with blue circles.} 
\label{fig:fov_example}
\end{figure}

\paragraph{Foveated Rendering}
In this section, we provide a comprehensive review of foveated rendering techniques and discuss gaze-based methods with similar objectives. In a nutshell, foveated rendering approaches focus on high-quality graphical computations and rendering around gaze fixation, while providing other parts of the scene in a relatively lower quality to reduce computational load without deteriorating the user experience. Figure~\ref{fig:fov_example} depicts image illustrations with blurred peripheral regions. 

To this end, \cite{towards_foveated_rendering_gazetracked_VR} proposed a foveated rendering architecture to lessen computational power and reduce the number of shades in the scene. This architecture generates a perceptual target image by adjusting the blur filter width according to retinal eccentricity to prevent temporal and spatial aliasing. The authors discovered that contrast enhancement could provide a larger tolerance for blurred region size through a user study. As a result, contrast enhancement was applied as a post-processing step to the image. The proposed technique achieved a similar level of temporal stability compared to temporally filtered non-foveated images. Another approach, rectangular mapping-based foveated rendering (RMFR), renders scenes with non-uniform foveation levels based on eccentricity and scene complexity~\cite{ye2022rectangular}. RMFR offers superior visual quality compared to conventional foveated rendering methods while requiring minimal rendering cost. \cite{3DKernel_Foveated_Rendering_for_Light_Fields_TVCG20} proposed 3D-kernel foveated rendering (3D-KFR), which is combined with an eye-tracking-based acceleration algorithm. The proposed algorithm, designed for visualizing high-resolution microscopic 4D light fields with depth cues, provides rendering acceleration up to a factor of $7.28$ in HMDs. 

Like the works above, \cite{deng2022fov} presented a gaze-contingent rendering approach with neural radiance fields (NeRF), which allows the rendering of 3D scenes with photo-realistic quality. However, this approach also requires heavy computation, which causes latency in the rendering process. To address this issue, the proposed technique involves encoding each retinal region, including the foveal, mid-periphery, and far-periphery regions, with varying levels of visual perception. Incorporating eye-gaze information reduces latency without compromising perceptual quality. Additionally, \cite{eye_domi_guided_fov_rendering_TVCG20} took eye dominance into account and designed an approach that provides more visual details to the dominant eye than to the non-dominant eye, thereby enhancing the efficiency of foveated rendering. An image with a broader foveation area is rendered for the eye with higher ocular dominance, and it showed a boost in the rendering performance of the devices using the proposed technique at the same level of perceived quality. 

Most discussed foveated rendering algorithms consider the peripheral region and the sensitivity of human eyes to model foveated areas. In addition, \cite{Luminance-Contrast-Aware_Foveated_Rendering_TOG19} introduced a content-aware foveated rendering method considering the luminance contrast of the exhibited content, which can forecast the resolution parameters as a function of luminance patterns by processing low-resolution frames before performing high-resolution rendering. Evaluations revealed that rendering performance increases while maintaining an imperceptible foveation layer. \cite{9583738} proposed a hybrid foveated rendering method incorporating foveation, field depth, and longitudinal chromatic aberration. This approach accounts for vergence and accommodation, visual acuity eccentricity, and color vision, and it outperforms the state-of-the-art rendering techniques in frame rates while providing at least the same level of visual quality. In parallel, \cite{9583734} presented a foveated rendering approach to address the high computational demands of ray tracing. The proposed approach incorporates a selective subsampling technique that gradually decreases the sampling rate in the peripheral region while maintaining high sampling rates in the foveal region. This approach allows the practical use of ray tracing in HMDs by reducing the high power requirement. 

A handful of rendering techniques rely on fixation prediction, and such techniques may suffer from latency, which can result in degraded user experience due to a mismatch between the actual gaze location and the predicted location, especially during saccades. To this end, \cite{Arabadzhiyska2017} provided an algorithm that forecasts the final position of saccades and then renders the image considering the predicted landing location. To determine the final saccade location, the direction of the movement is predicted based on the previous gaze samples, while the saccade amplitude is predicted using polynomial fitting, which can be integrated into existing foveated rendering systems to improve performance. Similarly, \cite{data_augmentation_for_saccade_landing_point_pred_etra20} proposed a data augmentation technique to enhance the saccade landing point estimation with neural networks. Time-shifted imitations of the training data are used to enhance the reliability of the estimation of the starting time of the saccadic movement. This augmentation improves the median accuracy of the predicted final saccade points for LSTMs and feed-forward neural networks. 

Apart from the discussed foveated rendering techniques, qualitative or quantitative performance evaluation approaches are necessary to properly assess and compare the performance of the existing rendering techniques. Additionally, the robustness and effectiveness of the introduced assessment methods may vary across different test scenarios. \cite{is_foveated_perceivable_in_VR_MM17} evaluated subjective assessment methods regarding efficiency and consistency for foveated rendering. Efficiency is assessed based on the required time to reach perceptual ratio convergence. Meanwhile, the distribution of individual quality of experience scores, which indicate user satisfaction, expectations, and perceptions, is used for consistency. The authors stated that no subjective evaluation method is superior to any other; however, they still provide the research community with information on evaluation metrics, allowing researchers to select a metric that best fits their specific needs. 

While we present various foveated rendering techniques designed to alleviate the high power and graphics processing unit (GPU) requirements of VR devices and make them practically usable for real-world applications, several other techniques based on eye tracking address similar challenges~\cite{duinkharjav2022color, focusVR_IMWUT18}. To this end, \cite{duinkharjav2022color} presented a real-time applicable power reduction method based on gaze and color discrimination, and achieved a display power requirement reduction of up to 24\% with minimal degradation in perceptual quality. \cite{focusVR_IMWUT18} proposed FocusVR to overcome the high power requirements of VR and AR devices with advanced high-resolution displays. They enhanced power consumption efficiency by combining techniques such as intelligent dimming, vignettes, and color mapping. The eye-tracking capability of these devices is utilized to execute more aggressive dimming techniques with less impact on user experience. FocusVR could reduce display and system power consumption up to 80\% and 50\%, respectively. In addition, \cite{refocusable_gp_panoramas_for_IVR_TVCG21} proposed an approach using eye tracking to address the high GPU requirements and lags associated with gigapixel panorama (GPP) displays in HMDs. A rendering technique supporting panning, tilting, zooming, and dynamic refocusing based on gaze was proposed to display GPP scenes in HMDs, and this technique can keep the frames per second (FPS) above 50 without a high-end GPU. 

\begin{figure}[h]
  \centering
  \includegraphics[width = 0.9\linewidth]{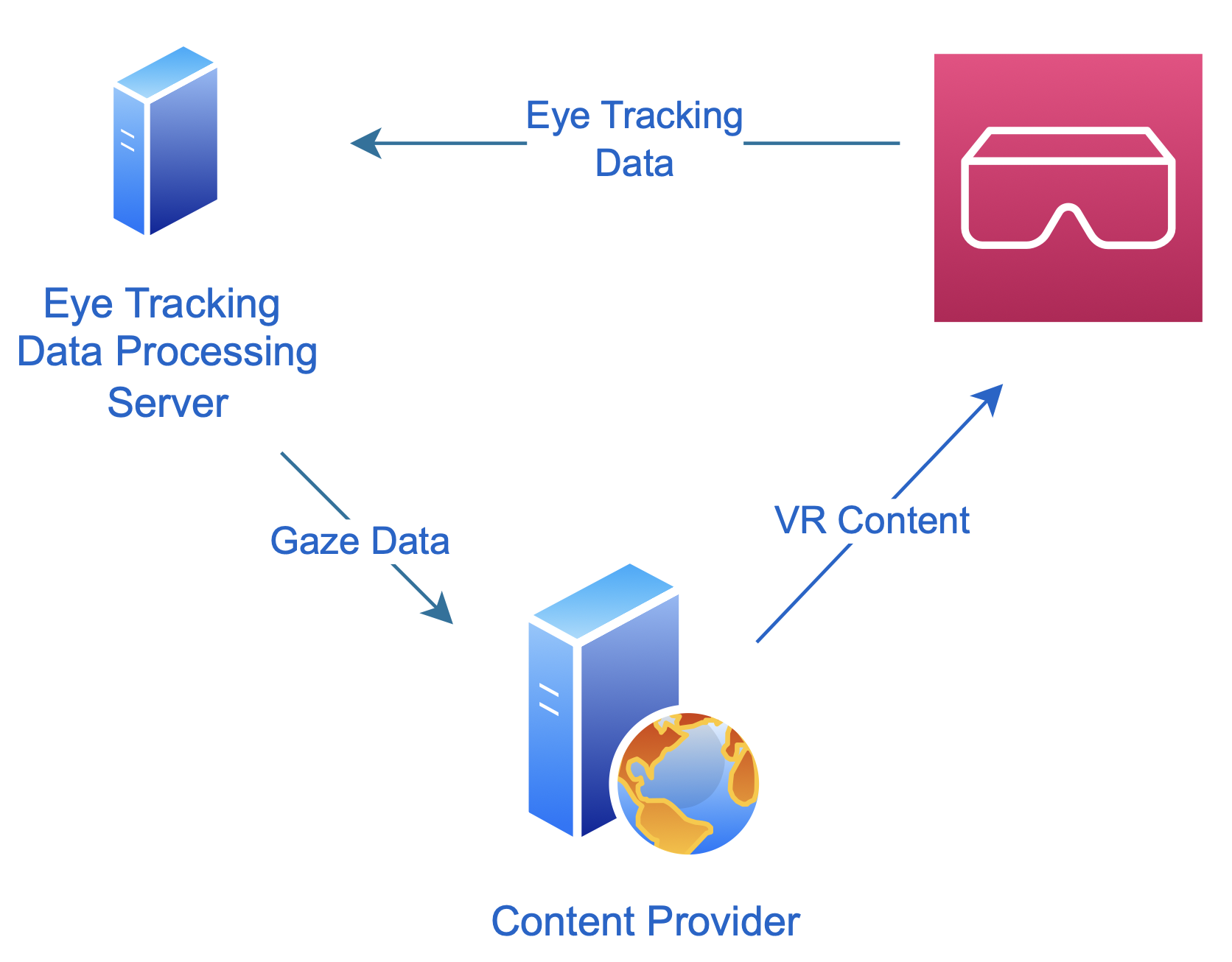}
  \caption{A streaming framework example. Eye-tracking data is provided to a server that estimates future gaze locations and communicates with a content provider. The content provider streams VR content considering gaze position and optimizes the transmission process.}
  \label{fig:streaming_1}
\end{figure}

\paragraph{Streaming Techniques}
Foveated rendering techniques can also be helpful or customized to reduce the bandwidth requirements of the high-resolution 360-degree video streaming, especially for VR~\cite{log-rectilinear_transformation_for_fov_360deg_vid_streaming_TVCG21,gaze-aware_streaming_nextgen_mobileVR_TVCG18}. In such setups, systems process eye-tracking data, and content providers transfer the data based on gaze position analogous to foveated rendering. Despite different approaches, Figure~\ref{fig:streaming_1} depicts the general framework for streaming solutions to optimize bandwidth efficiency. 

To this end, \cite{log-rectilinear_transformation_for_fov_360deg_vid_streaming_TVCG21} proposed a log-rectilinear transformation-based method using summed-area table filtering and off-the-shelf video codecs, that optimizes huge bandwidth requirements of the high-resolution live video streaming services for VR HMDs. Log-rectilinear-based transformation achieves approximately 10\% efficiency in bandwidth usage compared to the traditional log-polar transformation. Alternatively, \cite{gaze-aware_streaming_nextgen_mobileVR_TVCG18} presented a novel codec method that intentionally introduces errors in less-sensitive visual areas to support streaming video frames with regions of varying quality. This method offers an 83\% improvement in bandwidth requirement for high Quality of Service (QoS) levels compared to traditional solutions. Additionally, \cite{chen2022instant} proposed a gaze-contingent streaming approach that identifies the essential regions in 2D scenes and maps them to 3D scenes to enhance rendering and streaming efficiency. This approach is argued to provide better visual scene quality with temporal consistency compared to others. 

\paragraph{Improving Scene Realism and Depth Perception}
Foveated rendering or streaming techniques offer solutions to address or mitigate VR devices' technical and computational limitations, such as power consumption, computational load, and bandwidth requirements. In addition, achieving high scene quality and high levels of realism are other challenging concepts in VR, which are related to the visual acuity of scenes. Most research in this direction focuses on depth perception, which is associated with the degree of realism. Therefore, in the following, we discuss the rendering techniques introduced to enhance realism, with a primary focus on users' depth perception. We also explore the proposed solutions for improving depth perception, which addresses the vergence-accommodation conflict (VAC). 

\cite{gaze-contingent_depth_of_field_chi14} presented a gaze-contingent technique that allows for the production of realistic 3D images with improved depth perception by modifying the blurring rate of objects at varying distances. The authors achieved higher levels of perceived depth and realism by employing focal points in combination with the gaze. \cite{gazecontingent-ocular_parallax_rendering_VR_TOG20} proposed a rendering method based on ocular parallax exploiting small changes of images on the retina caused by eye rotation. The evaluations of ocular parallax's visibility for measuring perceived relative distance in AR and VR scenes indicated that ocular parallax rendering combined with gaze contingency offers a more authentic experience and improved depth perception in VR. 

Another work on ocular parallax that focuses on gaze-contingent stereo rendering that reduces shape and disparity distortions, providing a consistent sense of depth while maintaining perceived relative distances between the rendered objects in a scene~\cite{optimizing_depth_perception_VRAR_via_gaze-contingent_rendering_TOG20}. \cite{gazestereo3d_utilizes_fixation_gaze_est_TOG16} introduced another technique that utilizes eye tracking and predicted gaze data to enhance depth perception while preserving visual quality and user comfort. The proposed method employs a controller that executes unperceived local depth adjustments for stereoscopic content to optimize the trade-off between depth reproduction and visual comfort, significantly enhancing depth perception while preserving visual quality. 

VAC, the mismatch between the vergence and accommodation points of the eye, is another challenge that can affect the perceived image quality, the sense of depth, and realism in near-eye displays. This conflict arises due to changes in the shape of the lens of the eye during eye movements. In some VR displays, which are designed for a fixed distance, the degradation of perceived image quality can be attributed to the VAC. To address this, \cite{focal_surface_displays_TOG17} proposed the use of a phase-only spatial light modulator (SLM) made from the liquid crystal on silicon as a lens with a dynamically adaptive shape that is positioned between the display screen and the viewing optics. The authors presented a rendering technique that decomposes the scene and generates images for each display plane while considering the gaze location. 

Furthermore, \cite{fast_gaze-contingent_optimal_decompositions_TOG17} proposed a computationally plausible decomposition method that utilizes gaze and head motions to align the layers of the multifocal displays in HMDs. The evaluations indicated that eye tracking could be effectively utilized for plane alignment tasks. In addition, \cite{accomodation_invariant-comp_near-eye-display_TOG17} introduced an accommodation invariant display approach for near-eye displays, which makes scenes independent of the accommodation state of the eyes. This approach replaces traditional retinal blur with disparity and vergence to drive accommodation. \cite{batmaz2022effect} examined the effect of VAC on user performance during a hand-pointing task within arm's reach using multi-focal and single-focal VR displays, and concluded that multi-focal displays are better than single-focal displays in terms of both time and error rate, corresponding to the percentage of missed targets during virtual hand interaction tasks. More recently, \cite{aizenman2022statistics} systematically explored eye movement and binocular disparities when wearing an HMD and proposed an optimal screen distance to minimize the adverse effects of VAC by analyzing fixation distance distributions. However, eye-tracking accuracy can pose challenges for such solutions. To this end, \cite{dunn2019required} reported that the minimum required eye-tracking accuracy for varifocal displays in VR is $1.444^{\circ}$. 

\subsection{Understanding Human Cognition, Visual Attention, and Perception}
\label{sec_4_3}
Computer vision and machine learning methods for identifying eye regions and detecting gaze mostly work online for rendering or providing real-time user support. Researchers also often collect eye-tracking data from VR to analyze and understand human behaviors in an offline fashion. We first outline in Section~\ref{subsubsec:cognition} how eye tracking can aid in understanding human cognition in VR and during HMD use. Then, we explore the works investigating the relationship between human visual attention and gaze behaviors in Section~\ref{subsubsec:attention}. Later, in Section~\ref{subsubsec:perception}, we provide the research on human perception of VR, considering perceived realism and cybersickness. 

\subsubsection{Human Cognition} 
\label{subsubsec:cognition}
Unlike the tangible fields, the realm of human cognition is characterized by abstraction and is hence more elusive. To close this gap, eye-tracked VR has shown great potential in studying human cognition. Gaze features, such as eye movement patterns and pupil diameters, have been widely used as human cognitive process indicators. 

To this end, cognitive load prediction has been an important focus. \cite{bozkir_cognitive_load_vr19} studied cognitive load in driving simulations in VR and showed that it is possible to predict human cognitive load with an accuracy of over 80\% by incorporating eye features, particularly by using pupil diameters, with machine learning. \cite{souchet2022cognitive} conducted cognitive load prediction using features from different modalities, including gaze features, electrocardiogram, and electrodermal activity. The electrocardiogram features were the most relevant features correlated with self-reported subjective cognitive load levels compared to the other modalities. Furthermore, \cite{cognitive_load_task_diff_duchowski_chi18} pointed out the drawbacks of pupillometry in cognitive load assessment that are caused by ambient illumination and off-axis distortion and suggested the wavelet-based estimation of pupil diameter oscillation frequency as an alternative feature. The proposed metric demonstrated potential for classifying task difficulty while participants were performing math calculations. 

Researchers also utilized cognitive load prediction tasks in the aviation field. \cite{mental_workload_physiological_performance_features_ismar20} utilized features including pupil diameter size, pupil diameter amplitude, and constriction and dilatation speed of the pupil to evaluate the cognitive load in a VR flight simulation. Using random forests, the authors could classify four levels of cognitive load with an accuracy of up to 65\%. Furthermore, \cite{aviation_cognitive_load_IMWUT21} investigated cognitive load estimation in flight simulations, achieving 80\% accuracy using pupil dilation and blink features. 

Studying the correlation between pupil dilation and the difficulty index in Fitts' Law~\cite{fitts1954information}, which describes the relationship between the time, distance, and accuracy of an individual's movement in reaching a given target while considering the level of cognitive load or task difficulty, \cite{fittslaw_of_pupil_dilations_inHMD_ETRA19} indicated that tasks that demand higher cognitive load lead to a larger pupil diameter. In contrast, there was no significant relationship between pupil diameter and the Fitts' Law index in motor task complexity. 

Measuring pupil dilation is challenging due to its sensitivity to illumination~\cite{Eckert_etal_2022_etra}. To overcome this, \cite{pupillary_light_response_VR_HMDs_vrst18} attempted to exclude the impact of the light source on pupil dilation in 2D monitors and HMDs using three different models. A linear calibration model yielded prediction errors of 0.42 and 0.60 mm for pupil constriction and dilation in 2D, and an exponential model had an error of 0.27 mm in VR. Similarly, \cite{Eckert_etal_2022_etra} presented a method to correct for light-induced pupil diameter changes, and found that estimating luminance by using the weighted average of the fixation and background area provides the best performance. Related to its applications, \cite{gao2022evaluation} compared the effects of five VR locomotion methods on cognitive response, including arm swinging, dash, grappling, joystick, and teleportation, by employing gaze features, such as blink rates and fixation durations, to measure cognitive load and discovered that locomotion methods could significantly influence the cognitive states in VR. 

Human emotions, such as fear and frustration, also play a crucial role in the human cognitive state, and an increasing number of studies have been devoted to understanding the relationship between gaze behavior and emotions in VR. For instance, \cite{fear_of_sea_cognitive_state_VR_vrst18} used pupil diameter changes to measure the influence of the fear of the sea on working memory by designing n-back memory tests and found that pupil diameter changes noticeably in the underwater simulations. \cite{luong2022characterizing} investigated users' pupillometry responses along with pulmonary, electrodermal, and cardiac data to capture sensations of fear, frustration, and insight, where the latter refers to the moment when a user suddenly understands the solution to a puzzle or clue. The authors achieved an F1 score of over 71\% to predict these senses using classification models, including logistic regression and SVMs. Furthermore, \cite{Stoeve_etal_2022} employed machine learning by utilizing gaze, pupil diameter, and blink information to successfully estimate goalkeeper stress in VR, which is relatable to cognitive load and processes. 

Apart from estimating the cognitive load, inferences about human cognition have also been used as a supportive source of information in various application domains, especially to enhance the interaction experience in VR. \cite{context-aware_online_adapt_mr_interfaces_UIST19} utilized cognitive load, which was estimated by the index of pupil activity metric based on pupil dilations, to optimize and eliminate the manual adjustment process of the level of information displayed in MR glasses. This approach reduced the number of secondary task interactions by 36\%. \cite{Kuebler_Kasneci_Vintila_2017} utilized cognitive load as an indicative factor to detect hazard situations in a driving simulator. SVMs trained with wavelet transformation components extracted from pupil diameter data were used to detect hazardous situations. The results demonstrated that pupil dilation could help indicate hazard perception in driving simulations; however, it cannot be relied upon as a stand-alone detection mechanism due to the number of false positives. 

One of the most critical domains when considering applications of human cognition is medicine, as human cognitive processes are closely related to neurological and psychological diseases. In addition, since it is possible to manipulate and control experimental conditions very precisely in VR, researchers often employ VR and eye tracking together in this domain. For instance, to understand the response of children with autism spectrum disorder (ASD) to facial expressions, \cite{autism_respond_to_facial_expressions_desktop_based_but_good_TVCG13} conducted a user study and analyzed gaze features such as ROIs, fixation durations, blink rates, and pupil diameters, the latter being indicative of cognitive load, and found that the way individuals with ASD identify emotional faces is significantly different from their typically developing peers. 

In another work, \cite{autism-eyegaze-vr16} introduced a novel multi-modal adaptive social interaction in VR and suggested using gaze information for adaptive intervention. The authors used ROIs to adjust the face occlusion of virtual characters in real-time. In their emotion recognition experiments in the context of ASD, a performance increase of 3\% associated with the novel gaze-sensitive mechanism was observed. Similarly, \cite{kim2022vista} presented a VR-based interactive social skill training system (VISTA) that uses eye tracking, including ROIs and pupil features, to understand the characteristics of people with ASD. The authors found that the ASD group was highly engaged with VISTA while showing a dramatic difference from the neurotypical group in biological signals, such as a larger variation in pupil diameters, which indicates a higher cognitive load. 

Research in the medical domain, which includes human cognition, eye tracking, and VR, is not limited to the ASD context. \cite{parkinsons_disease_and_useful_for_diagnoses_TVCG17} presented a remote diagnosis system in VR for neurodegenerative diseases like Parkinson's. The authors designed various tasks to elicit abnormal eye behaviors such as ocular tremors, square wave jerks, and abnormal pursuits, which could indicate neurodegenerative diseases and cognitive load. The VR interface successfully elicited five types of abnormal eye movements, whereas eye physicians could identify three out of four abnormal eye behaviors. The authors also implemented different visualization techniques for eye movements, which can be helpful for doctors, as humans can hardly identify small movements. While the diagnosis of such diseases cannot be made solely with the help of eye tracking and VR, as these diseases are far more complex for machines and data-driven practices to identify, such systems can support doctors in their regular workflows. Table~\ref{tab:cognition} provides an overview of papers in Section~\ref{subsubsec:cognition}, including their characteristics, eye-tracking features, and the number of participants, whereas Table~\ref{tab:tab_short_names} includes abbreviations for the features. 

\begin{table}[!ht]
\caption{Feature abbreviations.}
\label{tab:tab_short_names}
\centering
\begin{adjustbox}{width=0.47\textwidth}
\begin{tabular}{ll|ll}
Name & Abbreviation & Name & Abbreviation\\
     \hline \hline
     Pupil-based features & pup & Blinks \& Eye openness & blnk \\
     Fixation & fix & Saccade & sac \\
     Interpupillary distance & ipd & Oscillation frequency & freq \\
     Eye vergence angle & eva & Eye gaze & gaze \\
     Phoria & phr & Convergence & cvgn \\ \hline
\end{tabular}
\end{adjustbox}
\end{table} 

\begin{table}[h!]
\scriptsize
\caption{An overview of the papers related to human cognition. ``\#P'' corresponds to the number of participants.}
\label{tab:cognition}
\centering
\begin{tabular}{llll}
Characteristics & \#P & Features \\ \hline \hline
\cite{bozkir_cognitive_load_vr19} Cognitive load prediction in driving & 16 & pup \\ 
\cite{souchet2022cognitive} Cognitive load prediction & 92 & pup, sac, blnk \\
\cite{cognitive_load_task_diff_duchowski_chi18} Cognitive load prediction & 17 & pup, freq \\
\cite{mental_workload_physiological_performance_features_ismar20} Cognitive load prediction in aviation & 75 & pup \\
\cite{aviation_cognitive_load_IMWUT21} Cognitive load prediction in aviation & 40 & pup, blnk \\
\cite{fittslaw_of_pupil_dilations_inHMD_ETRA19} Cognitive load analysis in physiology & 27 & pup \\
\cite{pupillary_light_response_VR_HMDs_vrst18} Cognitive load analysis & 24 & pup \\
\cite{Eckert_etal_2022_etra} Cognitive load analysis & 17 & pup \\
\cite{gao2022evaluation} Cognitive responses during locomotion & 15 & blnk, fix, sac \\
\cite{fear_of_sea_cognitive_state_VR_vrst18} Emotion recognition & 29 & pup \\
\cite{luong2022characterizing} Emotion recognition & 24 & pup \\
\cite{Stoeve_etal_2022} Stress classification & 30 & gaze, pup, blnk \\
\cite{context-aware_online_adapt_mr_interfaces_UIST19} Cognitive load prediction & 12 & pup \\
\cite{Kuebler_Kasneci_Vintila_2017} Cognition for hazard perception & 31 & pup \\ 
\cite{autism_respond_to_facial_expressions_desktop_based_but_good_TVCG13} Understanding ASD & 20 & gaze, fix, blnk, pup \\
\cite{autism-eyegaze-vr16} Adaptive intervention for ASD & 12 & gaze, pup, blnk, fix, sac \\
\cite{kim2022vista} Social skill training for ASD & 20 & gaze, pup \\ 
\cite{parkinsons_disease_and_useful_for_diagnoses_TVCG17} Parkinson's disease detection & 16 & gaze, pup \\ \hline
\end{tabular}
\end{table} 

\subsubsection{Visual Attention}
\label{subsubsec:attention}
This section provides a comprehensive overview of prior works that link eye-tracked VR and human visual attention. Human visual attention is related to where people look in a scene, primarily through their eye gaze, and often correlates with cognitive and perceptual abilities. Additionally, visual attention is influenced by the design of 3D scenes and stimulus-related factors in VR. 

To understand these concepts, researchers have studied various aspects of human visual attention. \cite{attention_guidance_panoramic_vids_ieeevr20} explored how peripheral flicker and central arrow stimuli differently impacted visual attention guidance in virtual panoramic videos. The authors reported that the participants preferred the central arrow stimuli over the peripheral flicker stimuli, perceiving the former as more rewarding. Additionally, they claimed that traditional attention mechanisms might not fully apply to panoramic videos. \cite{attention_orienting_VR_cues_sandvic_makingtask_ETRA19} investigated the effect of endogenous and exogenous cues to orient visual attention with eye tracking and found that such cue types reduce the users' reaction time as initially anticipated. \cite{liu2022adaptive} performed unordered tasks more efficiently over time by employing dynamic visual cues to guide users' visual attention towards specific substeps of the performed task. This method enabled users to determine the sequence of task completion, which subsequently influenced the presented set of cues, which were dynamically updated based on hand proximity and eye gaze. The authors stated that visual cues based on eye gaze improve task completion times more significantly than cues based on hand proximity. Studying attention guidance, \cite{attention_guidance_chi20_HiveFive} introduced HiveFive, a guidance technique to direct the users' attention to context-relevant points using swarm motion, which achieves lower response latency, resulting in a less adverse impact on immersion. 

Notifications are often considered a form of external stimulus that may interrupt the user during virtual experiences, and visual attention analysis could help to understand the impact of notifications and potentially provide a less disruptive VR experience. To this end, \cite{msg_notifications_in_vr_chi20} explored the perception of notifications by VR users when they received them in a message form from the real world. The authors provided design suggestions to mitigate the disruptive effect of these notifications, such as by measuring users' level of engagement using eye tracking to identify convenient times for notifications and positioning the message in the user's peripheral area. Similarly, \cite{chen2022predicting} proposed a technique that utilizes a deep learning model to process time series sensor data, including gaze angles and gaze shift rates, for predicting the optimal time for notification. The authors' deep learning model predicted a favorable time for notifications with 71\% precision, and further improvements were observed in relation to user activities and engagement. The assessment of visual attention is also considered useful for identifying the appropriate time and region to insert unperceived modifications within a virtual scene. Mise-Unseen~\cite{Mise-Unseen_UIST_19} is an approach that injects scene changes imperceptibly into the users' field of view. To adjust the injection time, the authors used eye tracking to assess the user's attention, intent, and spatial memory, demonstrating that gaze data in combination with masking techniques enables the insertion of modifications without being perceived by the user. 

\paragraph{Attention-based Virtual Space Design}
Visual attention analysis is useful for the design phase of virtual spaces. 3D virtual spaces could be designed to support the primary task objectives by analyzing and understanding visual attention. To this end, \cite{vis_att_optimizing_obj_placement_via_gaze} presented an attention-based approach to optimize the placement of visual elements and designed a 3D environment considering predefined goals. The authors employed a regression model to predict gaze duration and showed an optimization efficiency, combining cost functions for regularization, number of elements, and the primary design objective related to the visual elements to be placed. \cite{design_application_realtime_vis_model_3dVE_TVCG12} introduced an algorithm for exploring virtual scenes by simulating human visual attention. The proposed algorithm, which can also be applied during first-person exploration that involves walking and turning movements, utilizes a surface element-based representation instead of a mesh-based representation. This approach enables the optimization of scene quality to accelerate the rendering process. 

\paragraph{Joint Attention}
With VR, it is easier to assess joint attention of users, as with each HMD and user, it is possible to get precise gaze and head orientations in VR. As the information on visual attention also promotes interaction and collaboration, joint attention has been a focal point for some. \cite{eyegaze_headgaze_in_collaborative_games_ETRA19} analyzed the effect of eye-gaze- and head-gaze-based visual attention sharing in cooperative games. In such VR games, sharing eye-gaze led to higher subjective ratings of teamwork and shorter game duration than sharing head-gaze. Similarly, \cite{parallel_eyes_chi16} evaluated the effect of the shared egocentric videos with gaze locations in perspective sharing. The behavior changes and decisions of the participants who shared experiences were analyzed, considering parallel views. During the drawing activity, the researchers observed that individuals can develop behaviors to complement their partner's memory and decision-making. Furthermore, even in complex situations, individuals can create mechanisms to comprehend their physical embodiment and spatial relationship with their peers.

In the realm of medicine, impairments in joint attention can serve as an early indicator of ASD. Training and enhancing the joint attention mechanisms are essential in interventions for individuals, for instance, with ASD. \cite{autism_joint_attention_ieeevr18} attempted to address skills training for the joint attention of ASD individuals using customizable virtual humans (CVH). In an educational drum-playing scenario, evaluations revealed that the CVH enhanced participants' visual focus on relevant regions of the scene, albeit at the expense of increased reaction time. 

\paragraph{Visual Attention in Virtual Learning Spaces}
Eye tracking has also found applications in pedagogy, particularly in virtual learning spaces. Especially with the impact of the COVID-19 pandemic, remote and immersive learning spaces have gained popularity, particularly in the form of virtual classrooms. In such settings, the visual attention of students and teachers plays a crucial role in the learning processes. To facilitate this, \cite{ahuja_etal_chi21_classroom_twins} proposed a novel 3D digital twin classroom simulation that can be accessed through both VR HMDs and a web interface, incorporating a computer vision-based head- and eye-tracking system. Two cameras were deployed to track students and teachers, respectively, and gaze data were visualized in various forms, such as students' saliency maps on the board and intersections of instructors' gaze on student planes. A controlled study showed that this novel non-HMD classroom gaze system almost halved the gaze prediction error of its predecessors. In contrast, a reliability of 92.54\% was reported in an in-the-wild assessment.

Focusing more on the virtual simulations, \cite{Gao_bozkir_chi21} utilized several gaze features such as fixation durations, saccade durations, and pupil diameters to investigate the influence of various classroom manipulations on learners in VR, including different sitting positions, virtual avatar styles, and peer hand-raising behaviors. The authors found that such manipulations affect user attention and cognition differently, which could lead to different learning and engagement outcomes. \cite{Bozkir_stark_VR21} also studied the same manipulations in an eye-tracked VR classroom, focusing on virtual objects and attention times related to learning and engagement in the classroom. The authors demonstrated that learners seated in the front part of the virtual classroom paid more attention to the teacher and the board, whereas those seated in the back visually focused more on their peers. The authors also discovered that cartoon- and realistically styled avatars attracted different amounts of attention. Later, \cite{gao2022evaluating} explored various gaze features in a virtual lecture to scrutinize the effects of social interactions. The authors considered peer-learner hand-raisings as a cue of social interaction in VR and found that hand-raising animations and the number of peer learners who raise their hands affect students' attention in different ways. Analyses of attention and cognition toward such cues in virtual learning spaces are critical, as student self-concepts can be positively or negatively influenced by these behaviors. 

Complementary to the research above, \cite{distracted_students_edu_vr_ieeevr20} focused on the teacher's perspective during a lecture in VR and proposed six different gaze visualization techniques from the teacher's view to identify distracted students. These visualization techniques covered gaze ring, gaze disk, gaze arrow, gaze trail, gaze trail with arrows, and gaze heatmap, and the results showed that gaze trail visualizations outperformed the others in terms of popularity, and the application of 3D gaze heatmaps was problematic. In addition, \cite{Khokhar_Borst_2022} found in an educational VR setup that users perceive increased interactivity of the teacher agents as more natural and appropriate, providing insights on the design implications for virtual learning spaces. Table~\ref{tab:visual_attention} summarizes the papers related to visual attention. 

\begin{table}[h!]
\scriptsize
\caption{Overview of the papers related to human visual attention. ``\#P'' corresponds to the number of participants.}
\label{tab:visual_attention}
\centering
\begin{tabular}{llllll}
Characteristics & \#P & Features & \\ \hline \hline
\cite{attention_guidance_panoramic_vids_ieeevr20} Attention analysis in panaromic videos & 25 & gaze \\ 
\cite{attention_orienting_VR_cues_sandvic_makingtask_ETRA19} Attention guidance in sandwich preparation & 20 & gaze \\ 
\cite{liu2022adaptive} Attention guidance in unordered tasks & 15 & gaze \\
\cite{attention_guidance_chi20_HiveFive} Attention guidance & 20 & gaze \\ 
\cite{msg_notifications_in_vr_chi20} Attention analysis & 40 & gaze \\
\cite{chen2022predicting} Attention analysis & 20 & gaze \\
\cite{Mise-Unseen_UIST_19} Attention analysis for scene changes & 15 & gaze \\
\cite{vis_att_optimizing_obj_placement_via_gaze} Attention analysis for environment design & 23 & gaze \\
\cite{design_application_realtime_vis_model_3dVE_TVCG12} Attention analysis for environment design & 12 & gaze \\
\cite{eyegaze_headgaze_in_collaborative_games_ETRA19} Joint attention analysis in collaboration & 40 & gaze \\
\cite{parallel_eyes_chi16} Joint attention analysis in collaboration & 40 & gaze \\
\cite{autism_joint_attention_ieeevr18} Joint attention training for ASD & 10 & gaze \\
\cite{ahuja_etal_chi21_classroom_twins} Attention analysis in education & 13 & saliency, gaze\\
\cite{Gao_bozkir_chi21} Attention analysis in education & 288 & fix, sac, pup \\
\cite{Bozkir_stark_VR21} Attention analysis in education & 280 & gaze \\
\cite{gao2022evaluating} Attention analysis in education & 280 & pup, fix, sac, gaze \\
\cite{distracted_students_edu_vr_ieeevr20} Gaze visualization in education & 26 & gaze \\ \hline
\end{tabular}
\end{table} 

\subsubsection{Perception} 
\label{subsubsec:perception}
Human visual attention and cognition provide a lot of information on how humans experience virtual spaces; however, the research in these fields does not directly answer how humans perceive them. Combining engineering and human physiology, researchers have also focused on the perception of VR, utilizing eye movements. While some findings on human perception overlap with those on cognition and visual attention, it is essential to address these findings individually and tackle the issues collectively to design immersive and usable virtual spaces. In this section of the paper, we provide a detailed overview of the research that examines human perception of VR using eye tracking.

Depth perception is a well-studied topic for human perception in VR. For instance, \cite{3dpoint_acc_in_the_fovea_and_periphery_with_help_of_ET_IVR_TVGC21} investigated human depth perception during a 3D painting task in foveal and peripheral regions, revealing that task accuracy depended on the target's location rather than the starting point. Highly accurate 3D painting tasks can be accomplished in virtual spaces when targets are arm-reachable. Nonetheless, the authors observed that overestimation of depth, and this overestimation was more significant in the cases starting from the periphery and terminating in the foveal region. Furthermore, \cite{arefin2022estimating} revealed the relationship between perceptual depth level and eye vergence angle and interpupillary distance in VR, stating that these features could reveal changes in perceptual depth. 

Research on human perception also goes beyond depth perception. \cite{serrano_sitzmann_cognitive_event_segmentation_VR_vid_TOG17} systematically investigated the human perception of continuity in VR movies, which have different cinematographic requirements from conventional 2D movies. The authors utilized eye-tracking data in conjunction with the videos and revealed new metrics, such as scanpath errors, that describe visual attention, and provided design insights for VR content creation. Additionally, \cite{perception_of_gaze_of_vol_chars_ieeevr19} examined HMD users' perception capacity to comprehend the gaze direction of virtual volumetric characters in varying display resolutions, virtual character positions, head rotations, and gaze directions. The authors indicated that perceptual accuracy is considerably position-dependent and influenced by the direction of view. 

\paragraph{Understanding Perceptual Limits}
Identifying users' perceptual limits is essential for VR applications that consist of reorientation and repositioning, which often require discrete or continuous scene changes. However, inserting the scene changes imperceptibly during the experience is essential. Optimizing the amount and timing of these changes according to users' perceptual limits ensures that users' virtual experience is not disturbed by these changes. To this end, researchers have examined the perceptual limits and abilities of users, particularly to provide them with tailored VR solutions. For example, \cite{body_follows_eye_chi20} examined the effect of unobtrusive and prompt virtual object movements on users' body posture. The authors provided design insights while assessing users' perceptual limits for unnoticeable position changes of the virtual objects during eye blinks. \cite{In_the_blink_of_an_eye_Langbehn_TOG18} investigated conscious or unconscious blinks that suppress human visual perception, and demonstrated that users could not perceive translational changes of 4-9 cm and rotations of $2^{\circ}-5^{\circ}$. Furthermore, they also stated that a 50\% improvement can be observed in curvature gain, which is defined as $\frac{1}{r}$ where $r$ corresponds to the radius of a circular path in the real world while moving straight in VR. 

In addition to the blinks, \cite{keynavara_robert_effect_of_constant_camera_rot_etra20} evaluated the human limits during saccadic behaviors and what effects and manipulations are imperceptible during saccadic suppressions that occur during constant horizontal and vertical camera movements by utilizing a Bayesian procedure. The authors reported that sudden horizontal camera movements, which were additionally applied to the continuously moving virtual camera on the vertical axis, were less observable. In another work, \cite{degree_of_saccadic_supression_image_change_ETRA18} examined the visual sensitivity of VR users against changes in a 3D scene during saccadic suppression. Findings revealed that certain image transformations, such as rotations along the roll axis, were more noticeable by users during horizontal saccades. Similarly, \cite{saccadic_suppression_navigation_redirected_walking_TVCG15} proposed using saccadic suppression to include rotations and translations into the scene. The authors investigated the limitations of human perception during detected saccadic periods and found that users could not perceive positional changes up to 50 cm in the direction of their gaze and rotational modifications up to $5^{\circ}$. 

\paragraph{Perceived Realism of Virtual Spaces and Perceptual Considerations for Virtual vs Real Worlds} 
Sense of realism is related to the visual characteristics of the 3D scenes we explore, the ecological validity of the simulations, and how we perceive those 3D spaces physiologically. As human visual attention drives our eyes and we perceive scenes accordingly, eye-tracking analysis paves the way for understanding and estimating the realism level of virtual environments. This approach also enables the development of techniques to enhance the perceived realism of VR environments, taking into account the perceptual limitations of the human visual system. 

To this end, the effects of light sources on human vision, such as temporal eye adaptation, perceptual glare, visual acuity reduction, and scotopic color vision, can be used to enhance the realism level in virtual scenes. \cite{gaze_dependent_light_sim_in_VR_TVCG20} utilized gaze direction and pupil size as means of lighting effect adjustment to produce more realistic low-light scenes that are tailored to the users' vision. The authors revealed that the effect of the light in virtual scenes is highly subjective. Perception and reaction differences between virtual- and real-world tasks also provide insights into creating realistic virtual environments, in which gaze activity can be considered crucial for understanding human behavior in real and virtual worlds~\cite{eye-gaze-activity-in-crowds,collison-avoidance-gaze-vr19}. 

To design realistic crowd simulations, \cite{eye-gaze-activity-in-crowds} investigated gaze behavior during crowd walking activity. The authors analyzed the differences in gazing activity in real and virtual simulations while there were varying crowd levels during virtual walking simulations. As the crowd level increased, participants' eye movements became more restricted, scanning a narrower portion of the street. Furthermore, the authors noted that users directed their attention to the individuals in front of them and provided design recommendations, such as using a constant number of neighboring agents and positioning them by considering collision risks. In a different study, \cite{collison-avoidance-gaze-vr19} exploited gaze data to compare collision avoidance behaviors in the real world and VR environments. Findings revealed that collision avoidance behaviors are similar across different settings. A work investigating the existence of the stare-in-the-crowd effect, which refers to the tendency of individuals to detect and observe the gaze directed at them, validated this effect, as well as a negative correlation between the dwell time associated with this effect and social anxiety scores, while providing insights to achieve realistic interactive environments with virtual avatars~\cite{raimbaud2022stare}. Considering the similarities and differences between the real and virtual worlds, \cite{gupta2022total} examined autobiographical memory (AM), which is often considered a vital factor in human perception. The authors revealed the effect of AM on physiological cues like eye gaze, pupil diameter, and electrodermal activity. 

Avatars are essential components of virtual spaces and reality, serving as proxies for social interactions. They also have different impacts on perceived realism. To this end, eye-tracking methods to generate realistic avatars raised some attention. For example, \cite{evaluation_o_selfavatar_eyemovement_TVCG13} augmented a virtual self-avatar with eye movement animations to achieve a more comprehensive virtual embodiment and improve self-recognition. The authors compared representative animations of real eye movements with simulated eye movements without requiring eye-tracking hardware and reported that the use of eye movement animations increases the subjective sense of self-identification.

Considering the gaze behaviors of virtual agents, \cite{plausibility_avatar_gaze_and_env_TVCG17} evaluated the plausibility of virtual musical performance with changing environmental conditions and gaze-based attributes of audiences and musicians. The authors found that the gaze behavior of the virtual agents and distractions that do not comply with the nature of the environment had substantial effects. Furthermore, \cite{ma2022visual} recently presented a framework to generate self-avatars with facial expressions and assessed the psychological effects of realistic avatars. Participants found the facial expressions of cartoon-like avatars to be more controllable than those of realistic-looking avatars. Additionally, the authors observed that participants had a higher sense of body ownership in the first trial, regardless of the avatar type.  

Eyes, faces, and heads are essential components of virtual avatars, and a handful of works have explicitly focused on simulating those. For instance, \cite{head-eye-motion_generator_TVCG12} introduced a method, based on Gaussian mixture models and nonlinear dynamic canonical correlation analysis, and linear regression, to simulate head movement, eye gaze, and eyelid movement based on speech input. The results showed superior performance compared to state-of-the-art algorithms in generating head and eye motions. \cite{FaceVR_thies_TOG18} introduced FaceVR, an image-based face synthesis approach that includes an eye-tracking algorithm based on monocular videos, aiming to render more realistic outputs on stereo displays. \cite{Unmasking_communication_partners_aivr20} proposed a low-cost solution to construct a person's face wearing VR glasses, utilizing GANs to produce face images. Additionally, the authors created an eye-tracking method that gives cues on iris position and gaze direction. Similarly, \cite{realtime_3dfaceeye_capture_wearing_VR_headset_MM18} presented a CNN-based 3D face-eye reconstruction technique for constructing a personalized 3D avatar with eye movements. The proposed algorithm can construct personalized avatars with the VR users' facial expressions and eye animations. Another CNN-based technique~\cite{high-fidelity_facial_speech_anim_VR_TOG16} supports speech animation and emotional expressions, using mouth and eye-tracking cameras for HMDs. The proposed approach outperforms state-of-the-art techniques in animation fidelity without requiring individual calibration. Furthermore, \cite{face_synthesis_zhao_ieeev19} proposed a face reconstruction algorithm using a personalized 3D head model along with a colorization algorithm for near-infrared eye images without causing the red-eye effect, which is defined as color distortion in the irises. 

\paragraph{Cybersickness}
While VR offers numerous exciting opportunities, HMDs may lead to several types of physical discomfort, often referred to as VR sickness or cybersickness, which are closely related to human perception. Eye tracking has been frequently employed to understand the causes and severity of cybersickness, as well as to develop methods for mitigating or preventing it. For example, \cite{9583838} proposed a cybersickness severity prediction algorithm utilizing built-in HMD sensors, including eye and head trackers. This algorithm relies on deep fusion network models, and through the evaluations with a VR video game, the authors demonstrated the capability of their models by incorporating eye- and head-tracking features with video stimuli, and achieved an accuracy of up to 87.77\%. In another work, \cite{islam2022towards} introduced a deep fusion network based on Deep Temporal Convolutional Networks (DeepTCN) fusing physiological, head-tracking, and eye-tracking data. DeepTCN could predict cybersickness 60 seconds in advance with a 0.49 mean-squared error (MSE) on a scale between 0 and 10. Similarly, \cite{motion_sickness_prediction_3DConvNets_TVCG19} employed eye movement features to predict cybersickness severity along with disparity and optical flow maps that respectively represent depth and movement in the image. The authors' 3D CNN model achieved better precision rates when it included eye-tracking features compared to previous work~\cite{padmanaban2018towards}. 

Cybersickness may occur in any virtual experience, but is more common in motion-based ones due to mismatched real and virtual movements. For instance, virtual locomotion is one of the leading causes of VR sicknesses, such as nausea and dizziness, due to one-sided movement in the virtual world. According to \cite{virtual_locomotion_survey_TVCG20}, sickness occurring during locomotion is due to the visual-vestibular conflict, and FOV constraints are deployed to mitigate VR discomfort by restricting users' FOV. 
\cite{effect_fov_restrictor_sickness_ieeevr20} presented a gaze position-based foveated FOV restriction method to improve existing techniques that rely on head gaze to mitigate cybersickness. These existing techniques are limited to cases where the head and eye gaze are aligned. The authors demonstrated that their proposed method enables users to experience a wider visual scan field while maintaining the same level of VR sickness and noticeability. Furthermore, considering the relationship between cybersickness and locomotion techniques, \cite{Qian_etal_2018} found that the method utilizing eye movements for input modality yields high cybersickness due to the lack of head tracking. 

\paragraph{Perceptual Expertise Level}
As users with different expertise and skills tend to have varying visual perception patterns, the applications of eye-tracked VR have advanced significantly in the realm of expertise training and assessment. Driving training is of particular interest among the applications. For instance, \cite{driver_training_vr18} studied improper driving habits in a driving simulation with eye-tracking-enabled VR headsets and helped participants improve their driving skills by designing personalized training routes that considered their perceptual habits. Participants who were trained in a customized VR setup outperformed those trained by conventional methods, with respect to response time in emergency situations, training persistence, and an evaluation score based on inappropriate driving actions (e.g., not signaling before a turn). In another domain, \cite{classification_of_understanding_training_aivr19} utilized eye tracking to measure and classify English language understanding in VR using features such as pupil diameters and eye movements, and reached 75\% and 62\% prediction accuracies for easy and medium words, and for hard words, respectively. 

Industrial applications are also important, as they are used in everyday life. To this end, \cite{utilizing_vr_gaze_Track_for_AR_chi20} developed an elevator maintenance simulation in eye-tracked VR to facilitate industrial AR prototypes, which are challenging to build due to safety concerns in the real world. The authors recorded several types of behavioral data from users, including eye behaviors, and enabled gaze data visualization through scanpaths and heatmaps during training playback in VR. Based on their survey, the domain experts hold constructive opinions regarding eye tracking and gaze visualizations for industrial training. Additionally, \cite{gisler2021indicators} examined the relationship between training success and human behaviors in sanitary apprenticeship tasks. The authors combined gaze positions, head movements, and attention durations on the focused objects by using statistical summaries to predict the training success of users. Experiments in an industrial training task achieved a 10-20\% improvement in predicting users' training success compared to a baseline model. In addition, \cite{training_police_room_Jour_VR21} examined skill training for searching and tracking in virtual police rooms. The authors designed virtual experiments in which officers were tasked with searching and gathering evidence for an investigation, while also considering various criminal activities. Analyses of perceptual-cognitive skills in multiple training scenarios, taking saccadic eye movements into account, indicated that training users for visual search expertise in VR is possible. 

\paragraph{Vision Impairment, Ocular Examination, and Medical Perception}
Previous research has also utilized eye tracking and VR to study vision impairment and ocular examination, as vision problems lead to a perception of the world differently. Like other subfields discussed, in VR, it is possible to conduct highly controlled, programmed experiments compared to in-the-wild settings, allowing researchers to study such phenomena, which is why researchers have attempted to design ecologically valid simulations to study vision impairment issues in VR. For example, \cite{cataracts_ieeevr19} simulated cataract patients' vision in a virtual environment to better understand the perception of people with cataracts. The authors used eye tracking for these gaze-dependent symptoms because cataracts lead to special visual effects at the lens center and periphery. 

More related to ocular examination using eye tracking in VR, \cite{Kim_Son_Lee_Yun_Kwon_Lee_2016} transplanted the Developmental Eye Movement test to VR HMDs, which is a clinical eye test widely used to determine abnormalities in visual function and to assess ocular motor skills. Subsequently, \cite{Kim_Son_Lee_Kwon_2019} designed the King-Devick test, a standard measurement for assessing saccadic eye movement and dynamic visual acuity, using HMDs in both VR and AR. Furthermore, \cite{hotta2019vr} designed a gaze-based ocular examination for visual field defects in a virtual environment using fixation and saccadic features. Compared to conventional tests that take over 30 minutes, the proposed method can be conducted in five minutes while retaining sufficient accuracy and improving reliability. Additionally, medical VR applications can extend beyond setups that require HMDs. For instance, \cite{Kuebler_etal_2015} analyzed patients' eye and head movements with homonymous visual field defects (HVFD) in a virtual driving simulation. The virtual environment featured a VR cabin, and a head-mounted eye tracker tracked participants' eye movements. The authors' findings confirmed their hypothesis that a particular subset of HVFD patients could enhance their viewing behaviors by increasing visual scanning, thereby improving their driving skills. Similar to earlier sections, Table~\ref{tab:perception} presents an overview of the works on human perception. 

\begin{table}[h!]
\scriptsize
\caption{Overview of the papers related to human perception. ``\#P'' corresponds to the number of participants.}
\label{tab:perception}
\centering
\begin{tabular}{llp{1.6cm}}
Characteristics & \#P & Features \\ \hline \hline
\cite{3dpoint_acc_in_the_fovea_and_periphery_with_help_of_ET_IVR_TVGC21} Depth perception analysis in 3D painting & 18 & gaze \\ 
\cite{arefin2022estimating} Depth perception analysis & 24 & ipd, eva \\ 
\cite{serrano_sitzmann_cognitive_event_segmentation_VR_vid_TOG17} Continuity analysis in VR movies & 49 & gaze \\ 
\cite{perception_of_gaze_of_vol_chars_ieeevr19} Perceptual limit analysis & 37 & gaze \\ 
\cite{body_follows_eye_chi20} Perceptual limit analysis & 27 & blnk  \\ 
\cite{In_the_blink_of_an_eye_Langbehn_TOG18} Perceptual limit analysis for redirected walking & 32 & blnk \\ 
\cite{keynavara_robert_effect_of_constant_camera_rot_etra20} Perceptual limit analysis & 36 & sac \\ 
\cite{degree_of_saccadic_supression_image_change_ETRA18} Perceptual limit analysis & 10 & sac \\ 
\cite{saccadic_suppression_navigation_redirected_walking_TVCG15} Perceptual limit analysis & 13 & sac \\ 
\cite{gaze_dependent_light_sim_in_VR_TVCG20} Perceptual realism analysis in 3D scenes & 5 & gaze, pup \\ 
\cite{eye-gaze-activity-in-crowds} Perception analysis in crowd walking & 21 & gaze, fix \\ 
\cite{collison-avoidance-gaze-vr19} Perception analysis for collision avoidance & 17 & gaze, fix \\ 
\cite{raimbaud2022stare} Perception analysis in psychology & 30 & fix \\ 
\cite{gupta2022total} Perception analysis & 20 & gaze, pup, blink \\ 
\cite{evaluation_o_selfavatar_eyemovement_TVCG13} Perceived realism analysis for virtual avatars & 12 & gaze \\ 
\cite{plausibility_avatar_gaze_and_env_TVCG17} Perceived realism analysis & 20 & gaze \\ 
\cite{ma2022visual} Perceived realism analysis for virtual avatars & 18 & gaze \\ 
\cite{head-eye-motion_generator_TVCG12} Perceived realism analysis for virtual avatars & 20 & gaze \\ 
\cite{FaceVR_thies_TOG18} Face \& eye synthesis for virtual avatars & 18 & N/A \\ 
\cite{Unmasking_communication_partners_aivr20} Face \& eye synthesis for virtual avatars & 5 & N/A \\ 
\cite{realtime_3dfaceeye_capture_wearing_VR_headset_MM18} Face \& eye synthesis for virtual avatars & 7 & N/A \\
\cite{high-fidelity_facial_speech_anim_VR_TOG16} Face \& eye synthesis for virtual avatars & 310 & N/A \\ 
\cite{face_synthesis_zhao_ieeev19} Face \& eye synthesis for virtual avatars & 3 & N/A \\ 
\cite{9583838} VR sickness prediction & 30 & gaze, pup, cvgn \\ 
\cite{islam2022towards} VR sickness prediction & 30 & \makecell[l]{gaze, pup, cvgn,\\ blnk} \\ 
\cite{motion_sickness_prediction_3DConvNets_TVCG19} VR sickness prediction & 96 & saliency \\ 
\cite{effect_fov_restrictor_sickness_ieeevr20} VR sickness mitigation & 22 & gaze \\ 
\cite{driver_training_vr18} Skill training for driving & 50 & gaze \\ 
\cite{classification_of_understanding_training_aivr19} Skill training in education & 16 & pup, sac, fix \\ 
\cite{utilizing_vr_gaze_Track_for_AR_chi20} Skill training for maintenance & 12 &  \makecell[l]{fix, scanpath,\\ saliency} \\ 
\cite{gisler2021indicators} Skill training for sanitary apprentice & 48 & gaze \\ 
\cite{training_police_room_Jour_VR21} Skill training for police room search task & 54 & \makecell[l]{gaze, sac, fix, \\ entropy} \\ 
\cite{cataracts_ieeevr19} Vision simulation for patients with cataract & 21 & gaze, pup \\ 
\cite{Kim_Son_Lee_Yun_Kwon_Lee_2016} Ocular visual acuity assessment & 39 & \makecell[l]{eye dominance,\\ ipd, cvgn, phr} \\ 
\cite{Kim_Son_Lee_Kwon_2019} Ocular visual acuity assessment & 30 & sac \\ 
\cite{hotta2019vr} Ocular visual field analysis & 2 & sac, fix \\ 
\cite{Kuebler_etal_2015} Visual field analysis in driving & 14 & sac, fix, gaze \\ \hline
\end{tabular}
\end{table}

\subsection{Summary}
With eye tracking becoming pervasive, three main directions exist in the literature. The computer vision and machine learning communities address eye region segmentation, gaze estimation, and eye event classification tasks to facilitate accurate attention tracking in VR. Deep learning-based end-to-end approaches are successful when considering various algorithms and methods, particularly for eye region segmentation and gaze estimation. To facilitate the use of eye-tracking technology in an everyday manner in VR, various researchers and device vendors have recently embedded eye trackers in VR HMDs and provided the community with publicly available datasets to facilitate benchmarking and open science.

Once the eye regions and gaze directions are available, gaze-based interaction, including target selection and text entry through eye movements, predictive gaze analyses covering gaze and scanpath prediction, and gaze-contingent rendering, are the most prominent research directions. In most of these works, real-time working capability and eye movement signal quality are essential to achieve a good interaction experience. Therefore, if the gaze estimation and eye-tracker calibration work accurately, it is likely that methods relying on gaze provide a good usability and user experience. 

While real-time working capability and eye-tracking data quality are essential for gaze-based interaction in VR, eye movements have also been studied offline to gain a deeper understanding of human cognition, visual attention, and perception. As one can assess these human aspects in various domains in VR, the studies have a broader focus. Prominent domains include education and medicine, as there is a pressing need for online teaching and learning, especially following the COVID-19 pandemic. In these domains, it is possible to control environmental and stimulus factors with precision in VR, which is particularly essential for the medical domain. Regardless of the studied domains, fixation, saccade, blink, pupil diameter, and scanpath statistics are commonly used to assess human cognition, attention, and perception in VR. 

\section{Security and Privacy in Eye Tracking and Implications for Virtual Reality}
\label{sec_privacy}
It is possible to extract and infer useful information about users with eye-tracking data in VR; therefore, eye tracking is considered a powerful modality. We mainly considered works that utilize or are directly applicable to VR. However, virtual environments are relatively novel, and researchers have not evaluated possibilities with eye movements extensively in VR compared to real-world settings. \cite{Liebling2014} previously stated that, apart from person identification, eye movements can reveal many user attributes, such as gender, sexual preferences, body mass index, health status, or tasks, mainly in real-world settings. Considering that human eye movements depend on the stimulus and context, we first argue that as long as users encounter similar stimuli in virtual environments, comparable inferences will be possible, like in the real world, as some of them, such as gender and task predictions, can already be carried out accurately in VR~\cite{steil_dp_etra19, bozkir2021differential}. We foresee many other inference possibilities in everyday VR. 

To this end, \cite{Xu2022Metaverse} stated that while the existence of eye-tracking data is vital for utility tasks such as improving the efficiency of VR rendering, it also gives opportunities, especially for companies, for targeted advertising based on such unique characteristics and demographics of users. Considering these, as identification of users might help personalize VR, we first discuss authentication possibilities in Section~\ref{subsec_eyebased_authentication_VR}. However, if the purpose is not to authenticate users, carrying out the same or similar tasks may result in a privacy breach. The National Institute of Standards and Technology (NIST) defines the privacy breach as ``the loss of control, compromise, unauthorized disclosure, unauthorized acquisition, or any similar occurrence where a person other than an authorized user accesses or potentially accesses data or an authorized user accesses data for an other than authorized purpose''~\cite{NIST_privacy_breach}. According to the NIST's definition, access and action by authorized entities or data processing with consent does not constitute a privacy breach. However, with increased digitalization, adversarial entities may cause unintended privacy breaches. Considering this, previous research studied how to preserve data and user privacy. Section~\ref{subsec_privacypreserving_ET_VR} covers these topics and discusses methods for preserving privacy when eye tracking is the focus in VR. 

\subsection{Eye-based Authentication for VR}
\label{subsec_eyebased_authentication_VR}
Visual interaction and user assistance are two of the most prominent applications of behavioral data, including eye movements, particularly in real-time VR applications. Apart from those, such behavioral data have also been utilized for authentication purposes. One major challenge for behavioral authentication is that the accuracy of the methods is not very high, which can cause unpleasant user experiences. 

Several works in the literature studied eye and gaze-based authentication in VR. \cite{behavioral_biometric_VR_chi19} studied authentication in VR by utilizing body motion data, including head, hand, and eye movement data for different tasks such as pointing, grabbing, walking, and typing. The results indicate that while the best-achieved identification accuracies are near $60\%$, the performance drops as the user group size increases. However, they did not utilize any gaze features beyond the gaze ray. \cite{behavioral_biometrics_VR_chi21} studied user authentication based on VR spatial movements and reached an accuracy of up to $90\%$. In addition, another study \cite{liebers_VRST_2021} demonstrated that gazing behaviors and head orientation can authenticate users in VR with nearly perfect accuracy. However, the limitation of those studies is the small sample size. Therefore, it remains an open question as to how the utilized features perform when more users are available and how usable the proposed authentication methods would be. That is why more distinguishing and fine-grained features are needed for eye-based authentication. One could achieve this using high-sampling frequency eye trackers. 

When we consider eye-based authentication, there are two primary directions: iris-based and eye movement-based authentication. On the one hand, while iris textures are like visual fingerprints and help authenticate users with very high accuracies~\cite{XSong2020, nguyen_2024_acm_compsurveys}, and iris-based authentication can also be used in real-world scenarios such as border crossings~\cite{Daugman2009}. Due to privacy reasons, many commercial HMDs with eye trackers do not provide raw eye images to users or applications (e.g., HTC Vive Pro Eye~\cite{vrcompare_htcviveproeye}). This precaution can somewhat mitigate the problem; however, in the case of iris-textures utilized in VR, data protection issues should be handled carefully, possibly by using encryption~\cite{XSong2020}. On the other hand, even if the iris textures are not captured or are inaccessible, eye movements will still be provided, making it plausible to carry out authentication using them in the background over a certain period. While we do not conduct an extensive review of eye-based authentication, we provide essential studies related to implicit biometric authentication, particularly utilizing eye movements in VR, and discuss why these studies are relevant to the privacy aspects of such data. For more detailed information on authentication, and particularly eye-based authentication, we refer the reader to the prior survey papers~\cite{stephenson2022sok, katsini_secpriv_survey_chi20}. 

Task-independent authentication is possible using eye movements~\cite{Kinnunen:2010:TTP:1743666.1743712}. More related to VR, \cite{Eberz:2016:LLE:2957761.2904018} argued that one could use specific eye movement features for biometric authentication in affordable consumer-level devices during everyday tasks such as reading, writing, and web browsing. While the authors did not use a VR setup, their sampling rate of 50 Hz is even lower than the sampling rates of today's consumer-grade HMDs' eye trackers. In follow-up work, \cite{Eberz_28_blinks_later_CCS19} presented a continuous authentication system based on eye movement biometrics and stated that for eye movement-based authentication, a precise calibration should be carried out and the effects of light sensitivity and task dependence of eye movements should be considered when designing the authentication systems. In another work, \cite{Zhang:2018:CAU:3178157.3161410} proposed a continuous authentication scheme based on eye movements for VR HMDs, demonstrating the potential for HMD personalization by continuously authenticating the wearer in the background. Similarly, \cite{blinkey_VR_authentication_IMWUT19} used blinks and pupil sizes for user authentication in VR HMDs. 

While previous work demonstrates the plausibility of biometric authentication based on eye movements, most of these approaches suffer from usability or privacy issues in the context of VR, which is similar to behavioral authentication using spatial movements. Concerning usability, authentication models based on eye and gaze movements do not have as high accuracies as iris authentication, which could irritate HMD users if errors occur constantly. Furthermore, such models primarily rely on the temporal movement of the gaze, regardless of whether authentication is explicit or implicit, which results in longer authentication times compared to using single iris images. \cite{Lohr2022EyeKnow} stressed the similar implications and partly confirmed that biometric authentication performance using eye movements is comparable to four-digit PIN entry when eye-tracking signal quality is extremely high, along with high sampling frequencies (i.e., 1kHz). The authors stated that five seconds of eye movement data are needed to achieve such performance. \cite{measurement_time_presentation_size_bio_identification_noVR_etra21} also measured measurement time and presentation size of stimuli on biometric authentication, and they found that using a three-second timeframe maximized the biometric authentication success rates. While the authors did not use a VR setup for this, one can transfer the results to VR as well, as long as the quality of the eye-tracking data and presentation sizes are comparable. 

These findings suggest that while implicit eye-based authentication may not offer the same level of usability and security as PIN-based authentication, eye movements exhibit patterns that can help identify individuals to a certain extent. This indication may thus be more useful for personalization, where each user's data samples are accumulated over time while using VR HMDs and utilized for authentication-related settings such as Two-Factor Authentication (2FA). Since such an authentication process should run continuously in the background, longer processing times would not irritate users, as they would not be actively aware of the ongoing process. 

Several works could support this argumentation. For instance, \cite{implementation_of_eyemovement-driven_biometrics_VR_ETRA18} proposed a real-time capable architecture for eye movement-based authentication in VR and argued that eye-based biometrics would become a standard way of authentication for VR. \cite{NDSS20_OcuLock} explored the human visual system as a novel authentication method for VR HMDs and proposed OcuLock, which is resistant against impersonation and statistical attacks while maintaining a stable performance over two months, with an increasing error trend in longer timespan. Increased error rates in longer timespan pose another prominent challenge than the authentication accuracy itself, which requires further research. Furthermore, \cite{biometric_norm_techniques_eyemovements_TIFS19} assessed the effectiveness of biometric feature normalization techniques, including real and synthetic eye movement features, and found that the effectiveness of biometric normalization techniques on real-world data depends on feature intercorrelation. \cite{RepliCueAuth} used VR to evaluate the usability and security of real-world authentication systems, showing VR's potential as a testbed for authentication. 

Despite several disadvantages, such as the need for consistent tasks, high-sampling frequency eye movement data, and a decrease in authentication performance with a larger sample pool, previous work consistently showed that eye movement patterns represent personal identities, which is useful for authentication and personalization. However, if users do not want to take advantage of these possibilities and yet want to use gaze-based interactions and obtain assistive support, the linkage of eye movements to user identity and characteristics implies privacy risks that should be handled methodologically, meaning that personal patterns associated with eye movements should be either hidden in the data or handled in a privacy-preserving manner. At the same time, one should maintain high interaction utility and user experience. 

\subsection{Privacy-preserving Eye Tracking for VR}
\label{subsec_privacypreserving_ET_VR}
Research in eye-based authentication~\cite{katsini_secpriv_survey_chi20}, privacy considerations for eye tracking~\cite{Liebling2014, Kroeger2020}, and privacy risks of data collection in XR~\cite{some_privacy_risks_of_xr_chi21} demonstrate that when authentication is not preferred, it is advisable to preserve the privacy of the individuals and specific attributes of the individuals in the data, in addition to good data hygiene practices. Furthermore, \cite{SOK_data_privacy_in_VR} highlighted data privacy issues in VR, considering threat and defense models, and included eye-tracking sensors and data as discussion points. This section covers the methodological research on privacy protection of eye-tracking data. To provide users with privacy-aware eye-tracking solutions, privacy-preserving methods have been applied to sensor data in the form of eye images, aggregated data in the form of saliency maps and feature vectors, and sample-level gaze position data. Each method transforms or modifies data in various ways, creating a trade-off between data privacy and utility that varies by method and the intended application of the data. 

Recently, formal methods such as differential privacy~\cite{dwork_DP_only} have attracted the attention of VR and eye-tracking communities. Differential privacy is an established approach, applied across multiple domains, most prominently in the context of databases and aggregate survey data. It aims to conceal the existence of an individual's records in a particular database. This privacy mechanism is achieved by adding randomly generated noise, as specified by an $\epsilon$ parameter, to query outputs, ensuring that including a specific participant in the database does not significantly alter the queried function outputs. In $\epsilon$-differential privacy, the smaller the $\epsilon$ is, the more private the mechanism becomes. In the following, we provide the formal definition of $\epsilon$-differential privacy. 

\begin{definition}{\emph{$\epsilon$-Differential Privacy ($\epsilon$-DP)~\cite{dwork_DP_only}.}} 
\label{def:DiffPrivacy} 
A mechanism $M$ is considered $\epsilon$-differentially private for all databases $D_{1}$ and $D_{2}$ that differ at most in one element for all $S \subseteq Range(M)$ with;
\begin{equation} 
\Pr[M(D_{1}) \in S] \leq e^{\epsilon} \Pr[M(D_{2}) \in S].
\end{equation}
\end{definition}

The eye-tracking community initially utilized standard differential privacy mechanisms, namely Gaussian and Exponential mechanisms, on saliency heatmaps~\cite{liu_dp_etra19} and aggregated eye movement features that were collected from VR reading tasks~\cite{steil_dp_etra19}. These works showed the potential of formal methods for VR, even with $\epsilon$ values in the vicinity of $\epsilon=1$; however, they suffer from the correlations in the data that could jeopardize privacy, especially from temporal correlations, as independent noise sampling from the standard mechanisms allows adversaries to reconstruct signals that are very close to the original ones. Despite the correlation issue, both works showed that it is possible to maintain high accuracy in utility tasks while preserving privacy. 

Following up on these works, \cite{bozkir2021differential} addressed the temporal correlation issue by translating the data into difference signals and using the frequency domain to add the randomly generated noise. The authors noted that, apart from the correlation challenge, the privacy-utility trade-off is particularly important; they also indicated that, despite the challenges, finding an optimal trade-off is computationally feasible within the range $0<\epsilon<1$. Considering that eye trackers may already provide noisy sensor readings due to technical issues, finding a privacy-utility trade-off by adding more noise is an additional challenge. \cite{otus_tool_diffpriv_2022} adopted a local differential privacy approach to address the privacy-utility trade-off in a gaze trajectory generation task, arguing that their approach could effectively protect individual user privacy without significant utility loss. Yet, the noise required to establish a strong differential privacy guarantee is known to harm utility and may end up masking valuable insights from the processed data~\cite{kifer2011no}. To address this, the authors of the previous works that studied differential privacy aimed to calibrate the noise in a way that while privacy is achieved in the reasonable ranges (e.g., $\epsilon<1$), they tried to keep the utility tasks (e.g., document type classification) accurate and such tasks performed with accuracies double the amount of chance~\cite{bozkir2021differential}. Yet, preserving privacy in this manner is not trivial. 

Researchers have also explored alternative privacy guarantees beyond differential privacy to gain a deeper understanding of the range of privacy-utility trade-offs associated with eye-tracking data, specifically $k$-anonymity and $k,\gamma$-plausible deniability~\cite{k_anonymity_etra22,10049660}. These alternative guarantees target the privacy risk of re-identification, and their mathematical formulation is related to the ability to match released data to the original identity in a dataset. Alternative guarantees enable dataset owners to preserve privacy by reducing the risk of linking data to identities, thereby mitigating harms related to the inferences that can be made from eye tracking, while retaining data utility across various applications. First, $k$-anonymity is a definition of privacy in the context of re-identification from a dataset as follows~\cite{k_anonymity_1998}. 

\begin{definition} {\emph{$k$-anonymity.}}
\label{def:kAnonymity}
Given a person-specific dataset $D_{1}$, a de-identified dataset $D_{2}$ is $k$-anonymized by privacy process $\mathcal{P} : D_{1} \mapsto D_{2}$ if all released features $\Gamma_d = \mathcal{P}(\Gamma) \in D_{2}$ cannot be recognized as $\Gamma$ with probability $> \frac{1}{k}$.
\end{definition}

The privacy guarantee is interpreted with a lone privacy parameter $k$ linked to the upper bound on re-identification. Utilizing k-anonymity on eye feature data, \cite{k_anonymity_etra22} showed that person re-identification accuracies drop to chance levels. In contrast, the utility of a model trained on privacy-preserving data remains within reasonable ranges when the document-type classification is the utility task, similar to the works on differential privacy. However, the methods used to achieve $k$-anonymity typically depend on the duplication or generalization of data, which is not ideal when releasing datasets for research or statistical purposes. This issue led \cite{10049660,k_anonymity_etra22} to introduce the guarantee of $k,\gamma$-plausible deniability to eye-tracking data. Plausible deniability differs from $k$-anonymity in that it applies explicitly to synthetic data generated by a model. All synthetic data generated by the model is tested to ensure that it meets the following privacy guarantee before release. 

\begin{definition} {\emph{Plausible Deniability.}}
\label{def:pd}
For any dataset $D$ where $|D|\geq k$, and any record $y$ generated by a probabilistic generative model $\mathbf{M}$ such that $y = \mathbf{M}(d_1)$ for $d_i \in D$, it is stated that $y$ is releasable with ($k$,$\gamma$)-plausible deniability if there exist at least $k-1$ unique records $d_2,..., d_k \in D \setminus \{d_1\}$, such that

\begin{equation} 
\gamma^{-1} \leq \frac{Pr\{ y=\mathbf{M}(d_i)\}}{Pr\{ y=\mathbf{M}(d_j)\}} \leq \gamma 
\end{equation}
where $k \geq 1$ is an integer and $\gamma \geq 1$ is a real number. 
\end{definition}

When considering datasets of gaze samples, researchers found that different formal methods yielded practical privacy-utility trade-offs across various applications. Namely, practical trade-offs were achieved for the utility task of activity type recognition using both plausible deniability and a sample-based differential privacy method. In contrast, for a gaze prediction task, only $k$-anonymity produced a dataset with minimal errors when training a deep neural network. The takeaway from these works is that the recommended method often depends on the target application and whether privacy protections require differential privacy or guarantees specific to re-identification. 

In another work, \cite{kaleido_Usenix_21} adapted a differential privacy approach for providing real-time privacy control of $\epsilon$ for eye-tracking data based on location indistinguishability~\cite{andres2013geo}. The key difference between this approach and the other methods described earlier is the spatial privacy parameter, $r$. The authors proposed tuning the value of $r$ to objects currently in view of the user, which requires an object detection model to be run in parallel to the privacy noise method. The method was applied in real-time to an interactive gaze-controlled action game using webcam-based eye tracking and offline to a VR dataset of video viewing. Subjective feedback from participants indicated that the game was enjoyable with real-time privacy protection. Furthermore, while the game scores decrease in high-privacy settings (i.e., $\epsilon = 0.5$) compared to the no-privacy setting, the reduction is modest, with participants still tending to enjoy the game experience. Hence, the privacy-utility and, in this particular case, the privacy-interactivity trade-off were achieved by finding an empirical balance between the randomly generated noise to protect privacy and asking users whether they enjoyed the interaction experience with privacy protection. However, their real-time interaction utility evaluation was limited to 11 participants. In addition, while the suggested method effectively protects against re-identification, the VR dataset application was limited to predicting the users' visual correction prescription. Table~\ref{tab:datasets_takeaways} summarizes the existing formal privacy methods for eye-tracking data, corresponding guarantee, data type, and whether the privacy-utility trade-off was considered practical. 

\begin{table*}[ht]
\centering
\caption{Summary of formal privacy methods applied to eye movement data.}
\label{tab:datasets_takeaways}
\begin{adjustbox}{width=0.95\textwidth}
    \begin{tabular}[t]{p{3.4cm}llP{3.25cm}c} \hline
                    Mechanism & Guarantee & Data type & Utility task & Practical trade-off \\ \hline \hline
                    Laplace-DP~\cite{liu_dp_etra19} & $\epsilon$-DP& Fixation map & Saliency map generation & $\times$ \\ 
                    Gaussian-DP~\cite{liu_dp_etra19} & $\epsilon,\delta$-DP& Fixation map & Saliency map generation & $\checkmark$ \\ 
                    $k$-same-select \,\,sequence~\cite{k_anonymity_etra22} & $k$-anonymity & Features & Document type classification & $\checkmark$ \\ 
                    Marginals~\cite{k_anonymity_etra22} &$k$,$\gamma$-PD & Features & Document type classification & $\times$ \\ 
                    Exponential-DP~\cite{steil_dp_etra19} & $\epsilon$-DP& Features & Document type classification & $\times$ \\ 
                    DCFPA~\cite{bozkir2021differential} & $\epsilon$-DP& Features & Document type classification & $\checkmark$ \\ 
                    CFPA~\cite{bozkir2021differential} & $\epsilon$-DP& Features & Document type classification & $\checkmark$ \\ 
                    $k$-same-synth~\cite{10049660} & $k$-anonymity& Samples & Activity type classification & $\times$ \\ 
                    Event-synth-PD~\cite{10049660} & $k$,$\gamma$-PD& Samples & Activity type classification & $\checkmark$ \\ 
                    Kal$\epsilon$ido~\cite{kaleido_Usenix_21} & $\epsilon$-DP & Samples & Activity type classification & $\checkmark$ \\ 
                    $k$-same-synth~\cite{10049660} & $k$-anonymity & Samples & Gaze prediction & $\checkmark$ \\ 
                    Event-synth-PD~\cite{10049660} &$k$,$\gamma$-PD & Samples & Gaze prediction & $\times$ \\ 
                    Kal$\epsilon$ido~\cite{kaleido_Usenix_21} & $\epsilon$-DP& Samples & Gaze prediction & $\times$ \\ 
                    Kal$\epsilon$ido~\cite{kaleido_Usenix_21} & $\epsilon$-DP& Samples & Gaze-based web game & $\checkmark$ \\ 
                    Laplace-DP~\protect\cite{otus_tool_diffpriv_2022} & $\epsilon$-DP& Features & Gaze trajectory generation & $\times$ \\ \hline
    \end{tabular}
\end{adjustbox}
\end{table*}

As most works that provide formal privacy guarantees add considerable noise to the data, this negatively affects the performance of utility tasks. When real-time interaction is not required, and the utility is often limited to data mining, privacy protection through formal methods by adding noise is reasonable. However, especially in real-time applications, the amount of noise necessary for a privacy guarantee can significantly deteriorate the user experience. 

Considering this issue, researchers proposed solutions by emphasizing the importance of real-time and practical use, aiming to minimize the amount of noise introduced to real-time gaze data while also reducing the risk of re-identification. \cite{David-John_TVCG21} presented a privacy protection method by utilizing spatial and temporal downsampling of gaze positions with additive Gaussian noise when streaming the data to decrease the granularity of the eye-tracking data. The findings show that re-identification rates drop significantly when proposed methods are applied, without a formal guarantee, whereas the performance of gaze prediction as a utility task is minimally affected. In another work, \cite{fuhl2020reinforcement} proposed training reinforcement learning agents by maximizing rewards for utility tasks (e.g., expertise prediction, document-type classification) and by minimizing them for privacy tasks (e.g., gender detection, person identification) to balance the privacy and utility. This approach outperforms privacy protection by utilizing GANs and differentially private manipulation; however, it also operates probabilistically, meaning that it does not formally guarantee privacy and is most appropriate when specific risks and adversary models are known and not expected to change. 

In addition to the formal and probabilistic approaches, researchers also studied function-specific guarantees focused on gaze estimation. \cite{bozkir_ppge_etra20} proposed a cryptography-based method utilizing a randomized encoding-based framework in a three-party setup, where one party is identified as a server (e.g., a cloud instance) that trains machine learning models using sensitive eye movement data, and the other two parties provide the sensitive data in a masked way, protecting the data privacy. While this work is suitable for real-time interaction in any VR application, the utilized privacy framework is limited to two data-provider parties and requires efficient communication between the involved parties. More recently, \cite{elfares_fl_gaze_estimation_2022} proposed a federated learning approach for appearance-based gaze estimation in the wild using pseudo-gradient optimization. In federated learning~\cite{pmlr_v54_mcmahan17a}, machine learning models are trained in a decentralized manner, implying that sensitive data samples are not distributed but remain locally, thereby preserving data privacy. The authors demonstrated that their adaptive federated learning approach outperformed vanilla federated averaging~\cite{pmlr_v54_mcmahan17a} in both person-independent setups and most person-specific setups for the gaze estimation task. In follow-up work, \cite{elfares_Etal_2024_etra} combined federated learning and secure multi-party computation for appearance-based gaze estimation, demonstrating that enhanced privacy does not result in a decrease in gaze estimation accuracy. While neither of the works above utilized data collected from VR to train and evaluate their models~\cite{bozkir_ppge_etra20, elfares_fl_gaze_estimation_2022}, since VR HMDs can already be considered personal devices, federated learning frameworks and decentralized data processing fit well, and the authors' approaches are directly applicable in VR. The decentralization concept was recently also proposed in an anonymous eye-tracking data collection protocol for VR by eliminating the third parties for data processing and manipulation purposes~\cite{bozkir_blockchain_eyetracking_aivr20}. While the authors eliminated third-party entities by using blockchains and smart contracts, wallets associated with know-your-customer validation can still lead to accurate user identification to a certain extent. 

Apart from protecting the privacy of eye movements and gaze, a considerable amount of work focused on degrading iris authentication~\cite{degrading_iris_John_Etra19, John_pixel_noise_etra20}, obfuscation~\cite{iris_obfuscation_etra21}, and protecting the personal identities~\cite{rubber_sheet_model_eye_vid_etra20} when iris textures are involved. To this end, \cite{degrading_iris_John_Etra19} proposed an approach by utilizing an optical defocus in an eye-tracker setup, and found that such defocus causes errors in the range of calibration errors typical of eye trackers. More recently, \cite{John_sec_util_iris_auth_TVCG20} analyzed the security-utility trade-off for iris authentication using the optical defocus in the chin-rest setup and found similar results for degradation. The authors indicated that evaluating a similar setup in an immersive VR environment was in the scope of future work. These works primarily aimed to find a balance between the degradation of iris authentication and the accuracy of gaze tracking in the context of a privacy-utility trade-off.

In another work, \cite{John_pixel_noise_etra20} proposed adding pixel noise to break the iris signature to protect users from spoofing attacks, and argued that replacing up to $50\%$ of the pixels in the eye image while keeping the gaze estimation error less than $2.5^{\circ}$ is possible. In further studies, \cite{iris_obfuscation_etra21} found that an optimal method to remove the iris signature without impacting gaze estimation combines an edge-preserving filter with additive noise. While these works did not directly utilize eye-tracking data collected from VR setups, as VR with HMDs provides a more controlled environment for eye-tracking data collection, we foresee similar results from VR setups, provided the underlying gaze estimation approaches are the same. In the context of VR eye tracking, \cite{rubber_sheet_model_eye_vid_etra20} proposed replacing the iris texture regions with synthetic iris templates using a Rubber Sheet Model on the OpenEDS dataset~\cite{OpenEDS_Open_Eye_Dataset}, collected using VR HMDs, and found that such video manipulations do not degrade the semantic segmentation and pupil detection performance, similar to the previous findings~\cite{John_pixel_noise_etra20}. 

In summary, privacy-preserving eye-tracking methods can be categorized into two groups: one focuses on protecting sensitive gaze and eye movement information over time, including features aggregated from gaze movements, such as fixation durations or saccade rates, and the other focuses on iris obfuscation and degradation. Ultimately, the recommended method depends on the target application and whether the system is running in real-time or being applied offline to a dataset. We note that real-time methods should be tuned to the privacy and utility context of a given application, and it is often challenging to find a good trade-off if formal guarantees are necessary. We also recommend that usability and user experience be explicitly evaluated when new privacy approaches are proposed, as they are essential and cannot be sacrificed to achieve high levels of privacy or security. 

\subsection{Summary}
Researchers and practitioners utilize eye movements in VR not only for interaction and understanding human behavior, but also for person identification and authentication. The individuals' unique viewing behaviors demonstrate that eye movements can be utilized for authentication purposes to a certain extent, as the identification accuracies of such mechanisms are not as high as those of iris-based mechanisms. Despite that, such mechanisms are helpful for 2FA, particularly for in-app purchases or similar tasks in VR. The number of users in respective authentication databases and eye-tracker sampling frequencies are two of the most critical factors for an accurate eye movement-based authentication. Overall, despite the challenges, eye movements can be used for authentication purposes. Additionally, various personal characteristics and demographics can be identified using this data. Therefore, when authentication or identifying such characteristics is not the primary purpose, one should consider utilizing privacy-aware methods for eye-tracking data. Most prior works suggest applying differential privacy, utilizing cryptographic approaches, and decreasing the granularity of the collected eye movement data to preserve the individuals' and data privacy in VR. 

\section{Discussion and Future Directions}
\label{sec_discussion}
Taking into account the wide range of possibilities enabled by eye tracking, the state-of-the-art methods for preserving the privacy of such data, and the best practices recommended by the VR and privacy communities, a careful and detailed consideration of privacy is essential for eye tracking~\cite{Liebling2014, Kroeger2020}. We identify three main directions for the eye-tracking and privacy communities, and discuss these in the following. 

\subsection{Privacy in Social VR and Metaverse}
Eye tracking is heavily integrated into discussions regarding privacy in social VR due to the critical role of eye movements as a non-verbal cue during social interaction. Social interactions present special privacy considerations based on context, as expectations vary between interactions in public spaces and interactions with friends or family in private spaces~\cite{anonymity_vs_privacy_privacy_in_social_vr_vrst20}. Potential solutions to establish privacy in social interactions include modulating the degree of accuracy or resolution of eye movements being mapped to a virtual avatar. On the other hand, researchers have found that non-verbal communication, through eye movements and gestures, can help preserve privacy for users who wish to avoid vocal communication in social VR~\cite{maloney2020talking}, suggesting that users desire control over VR sensor data streams to protect privacy. Analogous methods exist for modulating privacy in telecommunications today, as users commonly turn off their webcam feed during video calls to preserve the privacy of not paying attention or of their background environment. Generating synthetic face and eye animations directly from audio has been proposed~\cite{voice2face}, which can help preserve privacy from raw eye movements, but is still prone to animation artifacts or incorrectly relaying users' social cues. Open areas where advancements in understanding privacy for social VR would be critical include environments geared towards children~\cite{maloney2020virtual}, individuals with behavioral conditions such as ASD~\cite{kim2022workplace}, and communities with privacy norms that differ from the general population~\cite{maloney2021social}. 

Recent use of the term ``Metaverse'' involves a broad vision of the future of immersive reality. While exact definitions vary, the broad futuristic visions include a connected and integrated usage of VR that spans many aspects of our daily lives, superseding the use of the Internet today. The privacy concerns introduced by such deeply connected and immersive VR are yet to be realized, but the long-term implications of capturing eye-tracking data of users in such environments are a critical line of future research. While unexplored in current VR platforms that include functionalities akin to a Metaverse, an example of potential privacy concerns for the future is captured by the short story ``The Schuhmacher''~\cite{kohno2022schuhmacher}. The story represents a fictional reality where persistent behavior tracking for advertising to customers is a two-way street that both benefits a shopkeeper's business and has negative impacts on his relationship and social status within his town. Privacy concerns due to eye tracking on a societal scale are largely unexplored. To further establish the privacy risks associated with eye tracking in naturalistic, real-world scenarios, eye-tracking datasets such as the Visual Experience Database~\cite{visualexperiencedatabase} can be examined. Such databases capture large-scale gaze data from typical daily activities, as opposed to experimental tasks that comprise the vast majority of research datasets. Characterizing how frequently privacy-sensitive scenarios arise due to eye tracking, and the magnitude of the privacy risk from the users' perspective, can lead to an understanding of how to anticipate long-term risks of the eye-tracked Metaverse. 

\subsection{Privacy vs Utility and Usability}
Computational methods that attempt to preserve the privacy of the users mostly achieve privacy by adding a certain amount of noise, as was the case for the works that utilize differential privacy~\cite{steil_dp_etra19, John_pixel_noise_etra20, bozkir2021differential}. However, as reported by these previous works, the privacy-utility trade-off should be taken care of. In differential privacy, it is especially challenging to find a good spot in terms of privacy-utility trade-off, considering the eye-tracking datasets, because the longer the signals are, the more noise should be added due to higher sensitivities that contribute to the noise. As eye-tracking data is already a noisy source of information, especially when utilized in real time, adding additional noise to preserve privacy may ruin the user experience. Li et al.~\cite{kaleido_Usenix_21} addressed this issue by obtaining information on how much users enjoyed the experience when they provided different privacy-utility levels during game play in a desktop setup. With the immersion provided by the VR setups, the user experience and enjoyment levels already change without any privacy provided in the first place, and it is an open question of how the privacy-utility trade-off transfers to immersive setups. In addition, even if the privacy-preserving solutions are used offline, such as private data mining, as high amounts of noise may destroy the patterns in the data, the eye-tracking community needs to find private data representations for different utility tasks (e.g., privacy-preserving scanpath comparison~\cite{ozdel_etal_2024_etra}) that have little effect on the utility tasks while preserving privacy. 

Privacy-preserving computations aside, providing such solutions in an adaptive and user-centric way is equally important. For instance, one user might be purely interested in a better user experience and utility, while the other one might prefer a privacy-enhanced experience during the use of VR environments. In addition, the particular use case context can also impact the privacy-utility trade-off for a user. For example, in real-time gaze-based interaction for a point-and-shoot game, user subjective experience and performance are critical to consider, in contrast to machine learning-based back-end use case contexts such as gaze-based activity recognition~\cite{wilson_privacy_2024}. To achieve this, user privacy behaviors should be known beforehand, and virtual environments can be designed and adapted accordingly. Such privacy aspects, including preferences and attitudes along with usability, have been studied for other wearable devices, including lifelogging cameras~\cite{hoyle_etal_2015_wearable_cams}, or AR glasses~\cite{gallardo_ar_attitudes_pets2023}. Despite this, these devices utilize real environments with limited to zero amount of vision augmentation. We argue that studying similar user behaviors for VR and HMDs with a focus on eye tracking will enable the deployment of user-adaptive privacy schemes for virtual spaces. Such adaptations should also align with privacy regulations since regulations differ based on the countries or regions~\cite{gdpr_vs_ccpa}. 

\subsection{Stimulus and Environmental Aspects for Privacy-preserving Eye Tracking in VR}
Privacy concerns from eye-tracking data primarily depend on what the user is looking at and the context of their environment when data is being recorded. The stimulus being viewed determines what private information is at risk. For example, eye movement behavior during a VR driving simulator may reveal the driver's age or their medical conditions, such as visual field loss~\cite{kasneci2014driving}. However, a driving task would have an extremely low chance of revealing sexual orientation, of which accurate classification requires viewing and revealing erotic stimuli~\cite{rieger2012eyes}. Beyond just stimulus, we also must consider the user's context, such as whether they are in the comfort of their own home viewing VR content alone, playing VR games online with friends, or using VR at work for remote collaboration in a professional setting. Prior research has demonstrated how users feel about sharing gaze data with some environments, for example, indicating that users are more likely to agree to share gaze data with medical government agencies but are not comfortable sharing with their employer for internal use~\cite{steil_dp_etra19}. Thus, practical privacy risks and user expectations for eye tracking depend on both stimulus and environmental context. 

We identify the need for research into understanding which types of stimuli or experiences can reveal certain types of private information (e.g., age, gender, ethnicity, sexual orientation, and identity), how much data is necessary for accurate classification, and how frequently those stimuli appear in typical VR use. There is a pressing need to further quantify privacy risk across environments and bridge the gap between real-world use of VR and findings from laboratory studies in controlled conditions. 

Additionally, understanding user expectations and anticipated societal norms across environments is critical to producing effective privacy-preserving systems specific to different domains. One future research direction is to view eye-tracking data through the lens of contextual integrity\,(CI)~\cite{nissenbaum2004privacy}. CI is a theoretical framework of privacy that introduces three main concepts: context, informational norms, and contextual purposes or values. Contexts capture distinct social spheres that arise naturally as part of society, including politics, religion, healthcare, or education. Informational norms are best characterized as a flow of data that society deems appropriate and consists of a sender, a recipient, a data subject, a data type, and finally, a transmission principle that provides the logic for when and how a data flow can occur. Contextual purposes capture the social value of a context, for example, by sharing gaze data with a medical provider, a doctor may be able to diagnose a condition that otherwise would have gone unnoticed, which has inherent value to the data subject and is understood within society. Focused research is required to transfer the theoretical components of CI into a practical privacy mechanism for eye tracking. Similar efforts have been attempted in other context-based computer science research~\cite{benthall2017contextual}. Establishing models and methods that define and enforce societal norms on eye tracking is a grand challenge for the future socio-technical landscape of VR and eye-tracking technology. 

\section{Conclusion}
\label{sec_conc}
This paper comprehensively covers eye-tracking research in VR and its implications for security and privacy, including authentication schemes and privacy-aware methods. To this end, we scanned the scientific literature from 2012 to 2022 across 35 major venues for VR, eye tracking, and privacy. We further provided and discussed three main research directions, with a primary focus on privacy, and highlighted the importance of privacy, as well as utility and usability.

\section*{Acknowledgments}
Parts of this work were conducted when EB and EK were with the University of Tübingen, and they acknowledge funding by the Deutsche Forschungsgemeinschaft (DFG, German Research Foundation) under Germany's Excellence Strategy - EXC number 2064/1 - Project number 390727645 and the DFG - Project number 491966293. The authors used Grammarly to improve readability. 

\bibliography{references}
\bibliographystyle{IEEEtran}

\end{document}